\documentclass[11pt]{article}
\usepackage[dvips]{graphics}
\usepackage{epsfig}
\usepackage{chicago}
\usepackage{amsmath}
\usepackage{amssymb}

\newcommand{\mC}{\mathfrak{C}}
\newcommand{\mD}{\mathfrak{D}}
\newcommand{\mE}{\mathfrak{E}}
\newcommand{\mF}{\mathfrak{F}}

\begin{document}
\title{Dynamical Synapses Enhance Neural Information Processing:
Gracefulness, Accuracy and Mobility}
\author{C. C. Alan Fung$^1$, K. Y. Michael Wong$^1$, He Wang $^2$ and Si Wu$^{3,4}$
\\
$^1$ Department of Physics, Hong Kong University of Science and Technology,\\ Hong Kong, China
\\$^2$ Department of Physics, Tsinghua University, Beijing, China
\\$^3$ Institute of Neuroscience, Chinese Academy of Sciences, China
\\$^4$ State Key Laboratory of Cognitive Neuroscience $\&$ Learning, \\ Beijing Normal
University, China}
\date{}
\maketitle

\begin{abstract}
Experimental data have revealed that neuronal connection efficacy exhibits two forms of short-term plasticity, namely, short-term depression
(STD) and short-term facilitation (STF). They have time constants residing between fast neural signaling and rapid learning, and may serve as
substrates for neural systems manipulating temporal information on relevant time scales. The present study investigates the impact of STD and
STF on the dynamics of continuous attractor neural networks (CANNs) and their potential roles in neural information processing.
We find that STD endows the network with slow-decaying plateau
behaviors---the network that is initially being stimulated to an active state decays to a silent state very slowly on the time scale of STD rather than
on the time scale of neural signaling. This provides a mechanism for neural systems to hold sensory memory easily and shut off persistent activities
gracefully. With STF, we find that the network can hold a memory trace of external inputs in the facilitated neuronal interactions, which
provides a way to stabilize the network response to noisy inputs, leading to improved accuracy in
population decoding. Furthermore, we find that
STD increases the mobility of the network states.  The increased mobility enhances the tracking performance of the network in response to
time-varying stimuli, leading to anticipative neural responses.
In general, we find
that STD and STP tend to have opposite effects on network dynamics and complementary computational advantages, suggesting that the brain may
employ a strategy of weighting them differentially depending on the computational purpose.
\end{abstract}


\section{Introduction}
Experimental data have consistently revealed that the neuronal connection weight, which models the efficacy of the firing of a pre-synaptic
neuron in modulating the state of a post-synaptic one, varies on short time scales, ranging from hundreds to thousands of milliseconds (see,
e.g., \cite{Stevens95,Markram96,Dobrunz97,Markram99}). This is called short-term plasticity (STP). Two types of STP, with opposite
effects on the connection efficacy, have been observed.  They are short-term depression (STD) and short-term facilitation (STF).  STD is
caused by the depletion of available resources when neurotransmitters are released from the axon terminal of the
pre-synaptic neuron during signal transmission \cite{Stevens95,Markram96,Dayan01}. On the other hand, STF is caused by the influx of calcium
into the presynaptic terminal after spike generation, which increases the probability of releasing neurotransmitters.

Computational studies on the impact of STP on network
dynamics have strongly suggested that STP can play many important roles in neural computation. For
instance, cortical neurons receive pre-synaptic signals with firing rates ranging from less than 1 to more than 200 Hertz.  It was suggested that STD
provides a dynamic gain control mechanism that allows equal fractional changes on rapidly and slowly firing afferents to produce post-synaptic
responses, realizing Weber's law \cite{Tsodyks97,Abbott97}.  Besides, computations can be performed in recurrent networks by population
spikes in response to external inputs, which are enabled through STD by recurrent connections \cite{Tsodyks00,Loebel02}.

Another role played by
synaptic depression was proposed by Levina, Herrmann, and Giesel (2007).  In neuronal systems, critical avalanches are believed to bring about optimal
computational capabilities, and are observed experimentally. Synaptic depression enables a feedback mechanism so that the system can be
maintained at a critical state, making the self-organized critical behavior robust \cite{Levina07}. Herding, a computational algorithm
reminiscent of the neuronal dynamics with synaptic depression, was recently found to have a similar effect on the complexity of information
processing \cite{Welling09}.  STP was also recently thought to play a role in the way a neuron estimates the membrane potential information of the
pre-synaptic neuron based on the spikes it receives \cite{Pfister10}.

Concerning the computational significance of STF, a recent work proposed an interesting idea for achieving working memory in the prefrontal
cortex \cite{Mongillo08}. The residual calcium of STF is used as a buffer to facilitate synaptic connections, so that inputs in a subsequent
delay period can be used to retrieve the information encoded by the facilitated synaptic connections. The STF-based memory mechanism has the advantage of
not having to rely on persistent neural firing during the time the working memory is functioning, and hence is energetically more
efficient.

From the computational point of view, the time scale of STP resides between fast neural signaling (in the order of milliseconds) and rapid
learning (in the order of minutes or above), which is the time scale of many important temporal processes occurring in our daily lives, such as
the passive holding of a temporal memory of objects coming into our visual field (the so-called iconic sensory memory), or the active use of
the memory trace of recent events for motion control. Thus, STP may serve as a substrate for neural systems manipulating temporal information
on the relevant time scales. STP has been observed in many parts of the cortex, and also exhibits large diversity in different cortical areas,
suggesting that the brain may employ a strategy of weighting STD and STF differently depending on the computational purpose.

In the present study, we explore the potential roles of STP in processing information derived from external stimuli, an issue of fundamental
importance yet inadequately investigated so far. For ease of exposition, we use continuous attractor neural
networks (CANNs) as our working model, but our main results are qualitatively applicable to general cases. CANNs are recurrent networks that
can hold a continuous family of localized active states~\cite{Amari77}. Neutral stability is a key property of CANNs, which enables neural
systems to update memory states easily and to track time-varying stimuli smoothly. CANNs have been successfully applied to describe the
generation of persistent neural activities \cite{Wang01}, the encoding of continuous stimuli such as the orientation, the head direction and
the spatial location of objects \cite{Ben-Yishai95,Zhang96,Samsonovich97}, and a framework for implementing efficient population
decoding~\cite{Deneve99}.

When STP is included in a CANN, the dynamics of the network is governed by two time scales. The time constant of STP is much slower than
that of neural signaling (100-1000 ms vs. 1-10 ms).  The interplay between the
fast and the slow dynamics causes the network to exhibit rich dynamical behaviors, laying the foundation for the neural system to implement
complicated functions.

In CANNs with STD, various intrinsic behaviors have been reported, including damped oscillations \cite{Tsodyks98}, periodic and
aperiodic dynamics \cite{Tsodyks98}, state hopping with transient population spikes \cite{Holcman06}, traveling fronts and pulses
\cite{Pinto01,Bressloff03,Folias04,Kilpatrick10}, breathers and pulse-emitting breathers \cite{Bressloff03,Folias04}, spiral waves
\cite{Kilpatrick09}, rotating bump states \cite{York09,Okada09}, and self-sustained non-periodic activities \cite{Stratton10}. Here, we focus on those network states
relevant to the processing of stimuli in CANNs, including static, moving and metastatic bumps \cite{Wu05,Fung10}. More significantly, we find
that with STD, the network state can display slow-decaying plateau behaviors, i.e., the network that is initially being stimulated to an active
state by a transient input decays to the silent state very slowly on the time scale of STD relaxation, rather than on the time scale of neural signaling. This is a
very interesting property. It implies that STD can provide a way for the neural system to maintain sensory memory for a duration
unachievable by the signaling of single neurons, and shut off the
network activity of sensory memory naturally. The latter has once been a challenging technical issue in the study of theoretical neuroscience \cite{Gutkin01}.

With STF, neuronal connections become strengthened during the presence of an external stimulus. This stimulus-specific facilitation lasts for a period on
the time scale of STF, and provides a way for the neural system to hold a memory trace of external inputs \cite{Mongillo08}. This information can be
used by the neural system for various computational tasks. To demonstrate this idea, we consider CANNs as a framework for
implementing population decoding \cite{Deneve99,Wu02}. In the presence of STF, the network response is determined not only by the instant input
value but also by the history of external inputs (the latter being mediated by the facilitated neuronal interactions). Therefore, temporal
fluctuations in external inputs can be largely averaged out, leading to improved decoding results.

In general, STD and STF tend to have opposite effects on network dynamics \cite{Torres07}. The former increases the mobility of network states, whereas the
latter increases their stability. Enhanced mobility and stability can contribute positively to different computational
tasks. Enhanced stability mediated by STF can improve the computational and behavioral stability of CANNs. To demonstrate that enhanced
mobility does have a positive role in information processing, we investigate a computational task in which the network tracks time-varying
stimuli. We find that STD increases the tracking speed of a CANN. Interestingly, for strong STD, the network state can even overtake the moving
stimulus, reminiscent of the anticipative responses of head-direction and place cells \cite{Blair95,Okeefe93,Romani10}.

The rest of the paper is organized as follows.  After introducing the models
and methods in Section 2, we discuss the intrinsic properties of CANNs in
the absence of external stimuli by studying their phase diagram in Section 3.
In Sections 4 to 6, we study the network behavior in the presence of various
stimuli.  In Section 4, we consider the after-effects of a transient stimulus,
and find that sensory memories can persist for a desirable duration and
then decay gracefully.  In Section 5, we consider the response of the network to a noisy
stimulus, and find that the accuracy in population decoding can be enhanced.
In Section 6, we consider the response of the network to a moving stimulus, and find that
the tracking performance is improved by the enhanced mobility of the network
states.  The paper ends with conclusions and discussions in Section 7.  Our preliminary
results on the effects of STD have been reported in Fung, Wong, Wang, and Wu (2010).

\section{Models and Methods}
We consider a one-dimensional continuous stimulus $x$ encoded by an ensemble of neurons. For example, the stimulus
may represent a moving direction, an orientation or a general continuous feature of objects extracted by the neural system. We consider the
case where the range of possible values of the stimulus is much larger than the range of neuronal interactions. We can thus effectively take
$x\in (-\infty,\infty)$ in our analysis. In simulations, however, we will set the stimulus range to be $-L/2<x\le L/2$, and have $N$ neurons uniformly distributed in the range obeying a periodic boundary
condition.

Let $u(x,t)$ be the current at time $t$ in
the neurons whose preferred stimulus is $x$. The
dynamics of $u(x,t)$ is determined by the
external input $I^{\rm ext}(x,t)$, the network
input from other neurons, and its own relaxation.
It is given by
\begin{equation}
    \tau_s\frac{\partial u(x,t)}{\partial t}
    =-u(x,t)+I^{\text{ext}}(x,t)
 +\rho\int^\infty_{-\infty} dx'J(x,x')p(x',t)\left[1+f(x',t)\right]r(x',t)
    ,
\label{eq:dyn}
\end{equation}
where $\tau_s$ is the synaptic time constant, which is typically of the order of 2 to 5 ms, and $\rho$ the neural density.  $r(x,t)$ is the firing
rate of neurons, which increases with the synaptic input, but saturates in the presence of global activity-dependent inhibition. A solvable
model that captures these features is given by Deneve, Latham, and Pouget (1999) and Wu, Amari, and Nakahara (2002)
\begin{equation}
r(x,t)=\frac{C_ru(x,t)^2}{1+k\rho\int_{-\infty}^{\infty} dx'u(x',t)^2},
\end{equation}
where $k$ is a positive constant controlling the strength of global inhibition and $C_r$
is a constant whose dimension is $r\left(x,t\right)u\left(x,t\right)^{-2}$. This type of global inhibition can be achieved by shunting
inhibition \cite{Heeger92,Hao09}.

$J(x,x')$ is the baseline neural interaction from $x'$ to $x$
when no STP exists. In our solvable model, we choose $J(x,x')$ to be of the
Gaussian form with an interaction range $a$, i.e.,
\begin{equation}
    J(x,x')=\frac{J_0}{a\sqrt{2\pi}}\exp\left[-\frac{(x-x')^2}{2a^2}\right],
\end{equation}
where $J_0$ is a constant. $J(x,x')$ is translationally invariant,
in the sense that it is a function of $x-x'$ rather than $x$ or $x'$.
This is the key to generating the neutral stability of CANNs.

The variable $p(x,t)$ represents the pre-synaptic STD effect, which
has the maximum value of 1 and decreases with the firing rate of the
neurons~\cite{Tsodyks98,Zucker02}. Its dynamics is given by
\begin{equation}
  \tau_d\frac{\partial p(x,t)}{\partial t} =1-p(x,t)-\tau_d\beta p(x,t)[1+f(x,t)]
r(x,t), \label{eq:dp}
\end{equation}
where $\tau_{d}$ is the time constant for synaptic depression,
and the parameter $\beta$ controls the depression effect due to neuronal firing.

The variable $f(x,t)$ represents the pre-synaptic
STF effect, which increases with the firing rate
of the neurons but saturates at a maximum value
$f_{\text{max}}$. Its dynamics is given by
\begin{equation}
\tau_f\frac{\partial f(x,t)}{\partial t}
=-f(x,t)+\tau_{f}\alpha\left[f_{\text{max}}-f(x,t)\right]
r(x,t), \label{eq:df}
\end{equation}
where $\tau_{f}$ is the time constant for synaptic facilitation,
and the parameter $\alpha$ controls the facilitation effect due to neuronal firing.

The dynamical equations (\ref{eq:dp}) and (\ref{eq:df}) are consistent with the phenomenological models of STD and STF fitted to experimental
data \cite{Tsodyks98} (see Appendix A). From Eqs.(\ref{eq:dp}) and (\ref{eq:df}), we can calculate the steady state values of
$p$ and $f$, which are
\begin{eqnarray}
p&=& \frac{1}{1+\tau_d\beta(1+f)r}, \\
f&=& \frac{f_\text{max} \tau_f\alpha
r}{1+\tau_f\alpha r}.\label{eq:f_steady}
\end{eqnarray}
Hence, in the high frequency limit, $\tau_f
\alpha r\gg 1$, we have $f\approx f_\text{max}$,
which can be regarded as a constant, and
$p=1/\left[1+\tau_d\beta(1+f_\text{max})r\right]$.
In this case, we only have to consider the effect
of STD. In the low frequency
limit, $\tau_d\beta (1+f)r\ll 1$, and so $p\approx 1$ and we only need to consider the
effect of STF.  Note, however, that the terms ``high frequency limit"
and ``low frequency limit" are used figuratively.  The actual
limits should depend on the other parameters mentioned above.

Our theoretical analysis of the network dynamics is based on the observations that: 1) the stationary
states of the network, as well as the profile of STP across all neurons, can be well approximated as Gaussian-shaped bumps; and 2) the state
change of the network, and hence the profile of STP, can be well described by distortions of the Gaussian bump in various forms. We
can therefore use a perturbation approach developed by Fung, Wong, and Wu (2010) to solve the network dynamics analytically.

It is instructive for us to first review the
network dynamics when no STP is included. This is
done by setting $\beta=0$ in Eq. (\ref{eq:dp})
and $\alpha=0$ in Eq. (\ref{eq:df}), so that
$p(x,t)=1$ and $f(x,t)=0$ for all $t$. In this
case, the network can support a continuous family
of stationary states when the global inhibition
is not too strong. These steady states are
\begin{eqnarray}
\tilde{u}(x|z)& = & u_{0}\exp\left[-\frac{(x-z)^{2}}{4a^{2}}\right],
\\
\tilde{r}(x|z) & = & r_{0}\exp\left[-\frac{(x-z)^{2}}{2a^{2}}\right],
\end{eqnarray}
where $u_{0}=[1+(1-k/k_{c})^{1/2}]\left(C_rJ_{0}\right)/(4ak\sqrt{\pi})$, $r_{0}=[1+(1-k/k_{c})^{1/2}]/(2ak\rho\sqrt{2\pi})$ and $k_{c}=\rho
\left(C_rJ_{0}\right)^{2}/(8a\sqrt{2\pi})$. These stationary states are translationally invariant among themselves and have the Gaussian shape with a free
parameter $z$ representing the peak position of the Gaussian bumps. They exist for $0<k<k_{c}$, and $k_{c}$ is the critical inhibition strength
above which only silent states with $u_0=0$ exist.

\begin{figure}[ht]\begin{center}
\includegraphics[width=12cm]{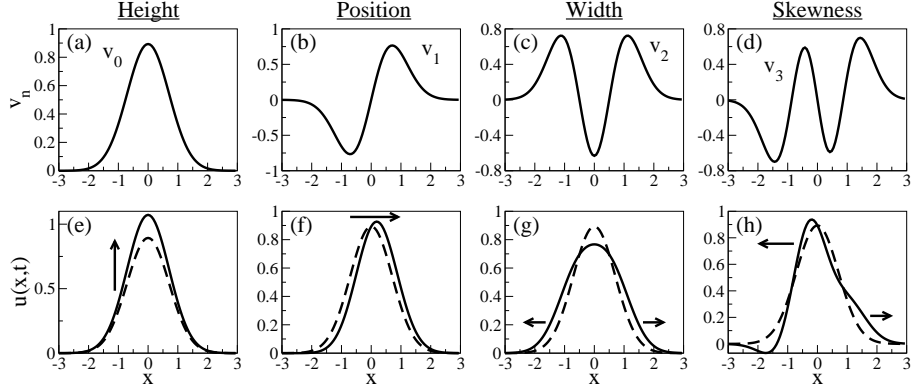}
\caption{(a) - (d) The first distortion modes of the bump
state.
(e) - (h) Their effects of producing distortions, respectively%
, in the height, position, width and skewness of the Gaussian bump.
Solid and dashed lines represent distorted and undistorted bumps
respectively.
\label{basis-functions}}
\end{center}\end{figure}

\begin{figure}[ht]\begin{center}
\includegraphics[width=10.3cm]{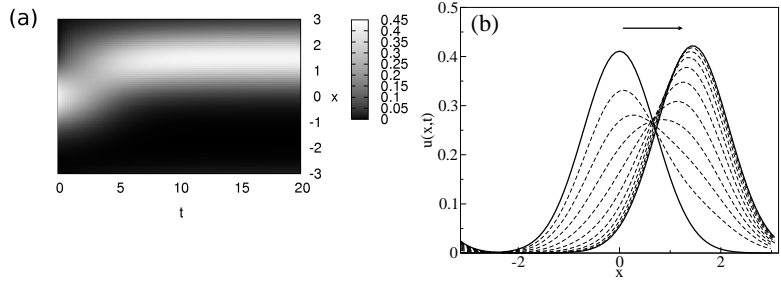}
\caption{(a) The neural response profile tracks the change in
position of the external stimulus from $z_0/a=0$ to $3$ at $t=0$. Parameters: $a = 0.5$, $\overline{k}=0.95$, $\overline\beta=0.0085$, $\rho C_r J_0
A=4.82843$. (b) The profile of $u(x,t)$ at $t/\tau=0,1,2,\cdots,10$ during the tracking process in (a). \label{tracking-example}}
\end{center}\end{figure}

Because of the translational invariance of the
neuronal interactions, the dynamics of CANNs
exhibits unique features. Fung, Wong, and Wu (2010) have shown that the wave functions of the
quantum harmonic oscillators can well describe
the different distortion modes of a bump state. For
instance, during the process of tracking an
external stimulus, the synaptic input $u(x,t)$
can be written as
\begin{equation}
  u(x,t)=\sum_{n=0}^\infty a_n(t)v_n(x|z(t)),
\label{eq:u_sum_u}
\end{equation}
where $v_n(x|z(t))$ are the wavefunctions of the
quantum harmonic oscillator
(Fig. \ref{basis-functions}),
\begin{equation}
  v_n(x|z) =  {\frac {(-1)^n(\sqrt{2}a)^{n-1/2}} {\sqrt{\pi^{1/2} n! 2^n}}}
    \exp\left[{\frac {(x-z)^2} {4a^2}}\right]
    \left(\frac{d}{dx}\right)^n
    \exp\left[-{\frac {(x-z)^2} {2a^2}}\right], ~~n=0,1,\ldots.
\label{eq:un}
\end{equation}
These functions have clear physical meanings,
corresponding to distortions in the height,
position, width, skewness and other higher order features of
the Gaussian bump (see
Fig. \ref{basis-functions}). We can use a
perturbation approach to solve the network
dynamics effectively, with each distortion mode
characterized by an eigenvalue determining its
rate of evolution in time. A key property of
CANNs is that the translational mode has a zero
eigenvalue, and all other distortion modes have
negative eigenvalues for $k<k_{c}$. This implies
that the Gaussian bumps are able to track changes
in the position of the external stimuli by
continuously shifting the position of the bumps,
with other distortion modes affecting the
tracking process only in the transients. An
example of the tracking process is shown in
Fig. \ref{tracking-example}, where we consider an external
stimulus with a Gaussian profile given by
\begin{equation}
I^\text{ext}(x,t)=A\exp\left[-\frac{(x-z(t))^2}{4a^2}\right].
\end{equation}
The stimulus is initially
centered at $z=0$, pinning the center of a
Gaussian neuronal response at the same position.
At time $t=0$, the stimulus shifts its center
from $z_0=0$ to $z_0=3a$ abruptly. The bump moves
towards the new stimulus position, and catches up
with the shift of the stimulus after a certain time,
referred to as the reaction time.

We can generalize the perturbation approach
developed by Fung, Wong, and Wu (2010) to study the dynamics
of CANNs with dynamical synapses. We will
only present the detailed analysis to the
case of STD.  Extension to the case of STF is
straightforward.

Similar to the synaptic input $u(x,t)$,
the profile of STD can be expanded in terms of the
distortion modes,
\begin{equation}
  p(x,t)=1-\sum_{n=0}^{\infty}b_n(t)w_n(x|z(t)),
\label{eq:p_sum_p}
\end{equation}
where $w_n(x|z)$ is given by
\begin{equation}
  w_n(x|z) = \frac{(-1)^na^{n-1/2}}{\sqrt{\pi^{1/2}n!2^n}}
  \exp{\left[\frac{(x-z)^2}{2a^2}\right]}\left(\frac{d}{dx}\right)^n
  \exp{\left[-\frac{(x-z)^2}{a^2}\right]}.
\end{equation}
Note that the width of $w_n\left(x|z\right)$ is $1/\sqrt{2}$ times that
of $v_n\left(x|z\right)$ due to the appearance of $r\left(x,t\right)
\propto u\left(x,t\right)^2$ in Eq. \eqref{eq:dp}.

Substituting Eqs. (\ref{eq:u_sum_u}) and (\ref{eq:p_sum_p}) into (\ref{eq:dyn}) and (\ref{eq:dp}), and using the orthonormality and
completeness of the distortion modes, we get the dynamical equations for the coefficients $a_n(t)$ and $b_n(t)$.  The details are presented in
Appendix B.

The peak position $z(t)$ of the bump is
determined from the self-consistent condition,
\begin{equation}
  z(t) = \frac{\int_{-\infty}^{\infty}dxu(x,t)x}{\int_{-\infty}^{\infty}dxu(x,t)}.
\end{equation}

Truncating the perturbation expansion at increasingly high
orders corresponds to the inclusion of
increasingly complex distortions, and hence
provides increasingly accurate descriptions of
the network dynamics.  As confirmed in the
subsequent sections, the perturbative approach
is in excellent agreement with simulation results.
The agreement is especially remarkable when the
STD strength is weak, and the lowest few orders
are already sufficient to explain the dynamical
features.  The agreement is less satisfactory
when STD is strong, and the perturbative approach
typically over-estimates the stability of the
moving bump. This is probably due to the
considerable distortion of the Gaussian profile
of the synaptic depression when STD is strong.

\section{Phase Diagrams of CANNs with STP}

We first study the impact of STP on the
stationary states of CANNs when no external input
is applied. For the convenience of analysis, we
will explore the effects of STD and STF
separately. This corresponds to the limits of high
or low neuronal firing frequencies, or the cases
where only one type of STP dynamics is significant.

\subsection{The Phase Diagram of CANNs with STD}

\begin{figure}[ht]\begin{center}
\includegraphics[width=8.3cm]{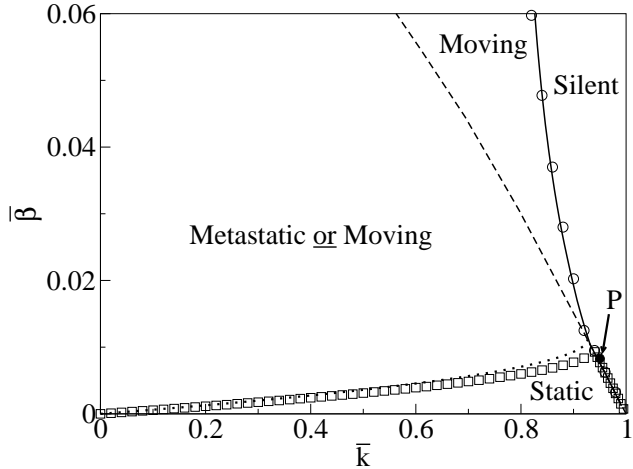}
\caption{Phase diagram of the network states with STD. Symbols: numerical solutions. Dashed line:
Eq.~(\ref{eq:static}). Dotted line: Eq.~(\ref{eq:static2}). Solid line: Gaussian approximation using the $11^{\rm th}$ order perturbation of the
STD coefficient. Point P: the working point for Figs. \ref{slow-decaying-behavior} and \ref{trajectory-theory}. Parameters: $\tau_d/\tau_s=50$,
$a = 0.5/6$, range of the network $x\in [-\pi, \pi)$.\label{phase-diagram-STD}}
\end{center}\end{figure}

We set $\alpha=0$ in Eq.~(\ref{eq:df}) to turn off STF. In the presence of STD, CANNs exhibit
new and interesting dynamical behaviors. Apart from the static bump state, the network also supports spontaneously moving bump states. Examining
the steady state solutions of Eqs. (\ref{eq:dyn}) and (\ref{eq:dp}), we find that $u_{0}$ has the same dimension as $\rho C_r J_{0}u_0^2$, and $1-p(x,t)$ scales as
$\tau_{d}\beta u_{0}^{2}$. Hence we introduce the dimensionless parameters $\overline{k}\equiv k/k_{c}$ and
$\overline{\beta}\equiv\tau_{d}\beta/(\rho^{2}\left(C_rJ_{0}\right)^{2})$. The phase diagram obtained by numerical solutions to the network dynamics is shown
in Fig. \ref{phase-diagram-STD}.

We first note that the synaptic depression and the global inhibition
play the same role in reducing the amplitude of the bump states.
This can be seen from the steady state solution of $u(x,t)$,
\begin{equation}
u(x)=\int dx'\frac{\rho J(x-x')u(x')^{2}}{1+k\rho\int dx''u(x'')^{2}+\tau_{d}\beta u(x')^{2}}.
\end{equation}
The third term in the denominator of the integrand arises from STD,
and plays the role of a local inhibition that is strongest where the
neurons are most active. Hence we see that the silent state with
$u(x,t)=0$ is the only stable state when either $\overline{k}$ or
$\overline{\beta}$ is large.

When STD is weak, the network behaves similarly to CANNs without STD, that is, the static bump state is present up to $\overline{k}$ near 1.
However, when $\overline{\beta}$ increases, a state with the bump spontaneously moving at a constant velocity comes into existence. Such moving
states have been predicted in CANNs \cite{York09,Kilpatrick10},
and may be associated with the travelling wave behaviors widely observed in the
neocortex~\cite{Wu08}. At an intermediate range of $\overline{\beta}$, the static and moving states coexist, and the final state of the
network depends on the initial condition. As $\overline{\beta}$ increases further, static bumps disappear. In the limit of high
$\overline{\beta}$, only the silent state is present. Below, we will use the perturbation approach to analyze the network dynamical behaviors.

\subsubsection{Zeroth Order: The Static Bump}

The zeroth order perturbation is applicable to
the solution of the static bump, since the
profile of the bump remains effectively Gaussian
in the presence of synaptic depression.  Hence,
when STD is weak and for $a\ll L$, we propose the
following Gaussian approximations,

\begin{eqnarray}
u(x,t) & = & u_{0}(t)\exp\left[-\frac{(x-z)^{2}}{4a^{2}}\right],\label{eq:U0_app0}\\
p(x,t) & = & 1-p_{0}(t)\exp\left[-\frac{(x-z)^{2}}{2a^{2}}\right].\label{eq:p_app0}
\end{eqnarray}
As derived in Appendix C, the dynamical equations for $u_0$ and $p_0$ are given by
\begin{eqnarray}
    \tau_{s}\frac{d\overline{u}(t)}{dt} & = &
    \frac{\overline{u}(t)^{2}}{\sqrt{2}(1+\overline{k}\overline{u}(t)^{2}/8)}
    \left[1-\sqrt{\frac{4}{7}}p_{0}(t)\right]-\overline{u}(t) + \overline{A},
    \nonumber \\
    & &
\label{eq:dU0_app}\\
    \tau_{d}\frac{dp_{0}(t)}{dt} & = &
    \frac{\overline{\beta}\overline{u}(t)^{2}}
    {1+\overline{k}\overline{u}(t)^{2}/8}
    \left[1-\sqrt{\frac{2}{3}}p_{0}(t)\right]-p_{0}(t),
\label{eq:dp0_app}
\end{eqnarray}
where $\overline{u}\equiv\rho C_rJ_0u_0$ is the dimensionless bump height and $\overline{A}\equiv\rho C_rJ_0 A$ is the dimensionless stimulus
strength.  For $\overline{A}=0$, the steady state solution of $\overline{u}$ and $p_{0}$ and its stability against fluctuations of
$\overline{u}$ and $p_{0}$ are described in Appendix C. We find that stable solutions exist when
\begin{equation}
    \overline{\beta}\leq
    \frac{p_{0}(1-\sqrt{4/7}p_{0})^{2}}{4(1-\sqrt{2/3}p_{0})}
    \left[1+\frac{\tau_{s}}{\tau_{d}(1-\sqrt{2/3}p_{0})}\right],
\label{eq:static}
\end{equation}
when $p_{0}$ is the steady state solution of Eqs.
(\ref{eq:dU0_app}) and (\ref{eq:dp0_app}). The
boundary of this region is shown as a dashed line
in Fig. \ref{phase-diagram-STD}. Unfortunately,
this line is not easily observed in numerical
solutions since the static bump is unstable
against fluctuations that are asymmetric with
respect to its central position. Although the
bump is stable against symmetric fluctuations,
asymmetric fluctuations can displace its position
and eventually convert it to a moving bump.  This
will be considered in the first-order
perturbation in the next subsection.

\subsubsection{First Order: The Moving Bump} 
\begin{figure}[ht]\begin{center}
\includegraphics[width=8.3cm]{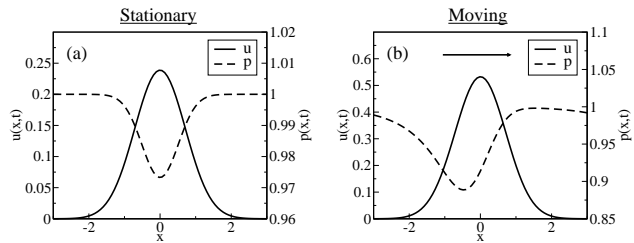}
\caption{ Neuronal input $u(x,t)$ and the STD coefficient $p(x,t)$ in (a) the static state at
$(\overline{k},\overline{\beta})=(0.9,0.005)$, and (b) the moving state at $(\overline{k},\overline{\beta})=(0.5,0.015)$. Parameter:
$\tau_d/\tau_s=50$.\label{phase-lag-STD}}
\end{center}\end{figure}

When the network bump is moving, the profile of STD is lagging behind due to its slow dynamics,
and this induces an asymmetric distortion in the profile of STD. Fig. \ref{phase-lag-STD} illustrates this behavior. Comparing the static and
moving bumps shown in Fig. \ref{phase-lag-STD}(a) and (b), one can see that the profile of a moving bump is characterized by the
synaptic depression lagging behind the moving bump.  This is because neurons tend to be less active in the locations of low values of $p(x,t)$, causing
the bump to move away from the locations of strong synaptic depression.  In turn, the region of synaptic depression tends to follow the bump.
However, if the time scale of synaptic depression is large, the recovery of the synaptic depressed region will be slowed down, and the region will be unable to catch up
with the bump motion.  Thus, the bump will start moving spontaneously.  This is the same cause attributed to anticipative non-local events modeled
in neural systems \cite{Blair95,Okeefe93,Romani10}.

To incorporate this asymmetry into the network
dynamics, we consider the first-order
perturbation. However, to facilitate our analysis,
we make a further simplification as follows:
\begin{eqnarray}
  u(x,t) &=& u_0(t)\exp\left[-\frac{(x-vt)^2}{4a^2}\right], \label{eq:moving_ansatz1}\\
  p(x,t) &=& 1-p_0(t)\exp\left[-\frac{(x-vt)^2}{2a^2}\right]  +p_1(t)\exp\left[-\frac{(x-vt)^2}{2a^2}\right]\left(\frac{x-vt}{a}\right).  \label{eq:moving_ansatz2}
\end{eqnarray}
This means that we have restricted the bump profile to the
zeroth order. Comparison with the full first-order perturbation shows that
the discrepancy is not significant.  This is because the
synaptic interactions among the neurons effectively maintain the bump
profile in a Gaussian shape, whereas the STD profile is
much more susceptible to asymmetric perturbations.

As described in Appendix D, we obtain four steady-state equations for $\overline{u}/B$, $p_0$, $p_1$ and $v\tau_s/a$ in terms of the
parameters $\overline{\beta}\overline{u}^2/B$ and $\tau_s/\tau_d$, where $B\equiv 1+\overline{k}\overline{u}^2/8$ is the global inhibition
factor.  It is easy to first check if the static bump obtained in Appendix D is also a valid solution by setting $v$ and $p_1$ to
0. We can then study the stability of the static bump against asymmetric fluctuations. This is done by introducing small values of $p_1$ and
$v\tau_s/a$ into the static bump solution and considering how they evolve.  As shown in Appendix D the static bump becomes unstable
when
\begin{equation}
  \frac{\overline{\beta}\overline{u}^2}{1+\overline{k}\overline{u}^2/8} \leq
  Q\left[\frac{\tau_d}{\tau_s} - R + \sqrt{\left(\frac{\tau_d}{\tau_s} - R\right)^2-S}\right]^{-1},
\label{eq:static2}
\end{equation}
where $Q=7\sqrt{7}/4$, $R=(7/4)[(5/2)\sqrt{7/6}-1]$,
and $S = (343/36)(1-\sqrt{6/7})$.  This means that in
the region bounded by Eqs. (\ref{eq:static})
and (\ref{eq:static2}), the static bump is unstable to
asymmetric fluctuations.  It is stable (or more precisely,
metastable) when it is static,
but once it is pushed to one side, it will continue to
move along that direction.  We call this behavior {\it metastatic}.
As we shall see,
this metastatic behavior is the cause of the enhanced
tracking performance.

Next, we consider solutions with non-vanishing
$p_1$ and $v$.  We find that real solutions exist
only if condition (\ref{eq:static2}) is
satisfied. This means that as soon as the static
bump becomes unstable, the moving bump comes into
existence. As shown in
Fig. \ref{phase-diagram-STD}, the boundary of this
region effectively coincides with the numerical
solution of the line separating the static and
moving phases.  In the entire region bounded by
Eqs. (\ref{eq:static}) and (\ref{eq:static2}),
the moving and (meta)static bumps coexist.

We also find that when $\tau_d/\tau_s$ increases,
the moving phase expands at the expense of the
(pure) static phase.  This is because the
recovery of the synaptic depressed region becomes
increasingly slow, making it harder for the region to catch up
with the changes in the bump motion, hence sustaining
the bump motion.

\subsection{The Phase Diagram of CANNs with STF}

\begin{figure}[ht]\begin{center}
\includegraphics[width=8.3cm]{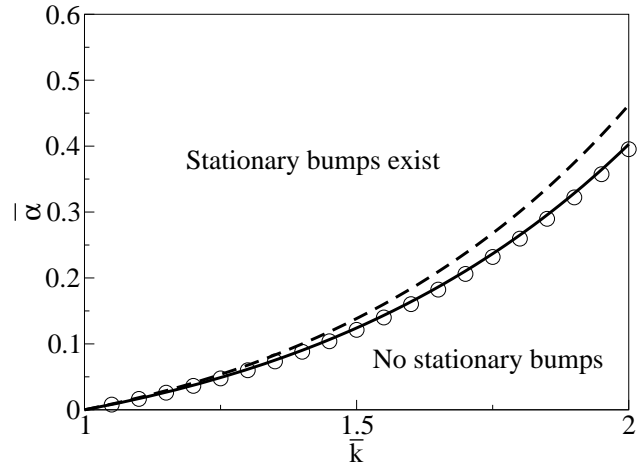}
\caption{Phase diagram of CANNs in the presence of STF.  As the synaptic facilitation is present, the range allowed to support
a stationary bump is broadened.  Dashed line: Prediction by $0^\text{th}$ order approximation.  Solid line:
Prediction by $2^\text{nd}$ order approximation. Symbols: Simulations. Parameters: $N/L = 80/(2\pi)$, $a/L = 0.5/(2\pi)$, $\tau_f/\tau_s = 50$
and $f_\text{max} = 1$. \label{phase-diagram-STF}}
\end{center}\end{figure}

We set $\beta=0$ in Eq. (\ref{eq:dp}) to turn off STD. Compared with STD, STF has qualitatively the opposite effect on the network dynamics.
When an external perturbation is applied, the dynamical synapses will not push the neural bump away.  Instead they will try to pull the bump back
to its original position.  The phase diagram in the space of $\overline{k}$ and the rescaled STF parameter $\overline{\alpha} \equiv \tau_f
\alpha / (\rho^2\left(C_rJ_0\right)^2)$ is shown in Fig. \ref{phase-diagram-STF}.  When $\overline{\alpha}$ increases, the range of inhibitory strength
$\overline{k}$ allowing for a bump state is enlarged.  Note that since STF tends to stabilize the bump states against asymmetric fluctuations,
no moving bumps exist. The phase boundary of the static bump is well predicted by the second-order perturbation.

Concerning the time scale of neural information
processing, it should be noted that it takes time
of the order of $\tau_f$
for neuronal interactions to be fully
facilitated. In the parameter range of
$\overline{k}>1$ where the facilitated neuronal
interaction is necessary for holding a bump
state, we need to present an external input
for a time up to the order of
$\tau_f$ before a bump state can be sustained.


\section{Memories with Graceful Degradation in CANNs with STD}

We consider the response of the network to a
transient stimulus given by
\begin{equation}
I^\text{ext}\left(x,t\right)
=
A\left(t\right)
\exp \left[ - \frac{\left(x-z_0\right)^2}{4a^2} \right].
\end{equation}
Here $A\left(t\right)$ is non-zero for some duration before $t=0$,
so that a bump is rapidly formed, but $A\left(t\right)$ vanishes after
$t=0$.

\begin{figure}[ht]\begin{center}
\includegraphics[width=8.3cm]{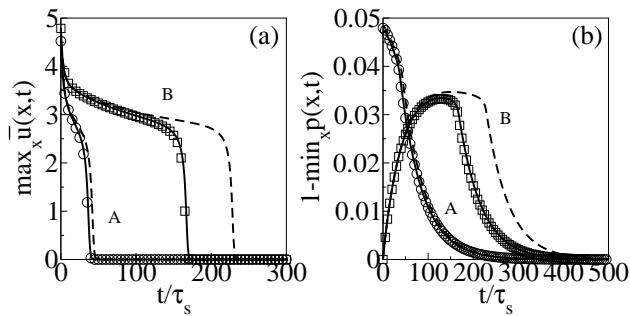}
\caption{ The height of the bump decays over time for two initial conditions of types A and B in
Fig. \ref{trajectory-theory} at $(\overline{k},\overline{\beta}) = (0.95,0.0085)$ (point P in Fig. \ref{phase-diagram-STD}).
 Symbols: numerical solutions. Dashed Lines: $1^\text{st}$ order
perturbation using Eqs.~(\ref{eq:dU0_app}) and (\ref{eq:dp0_app}). Solid lines: $2^\text{nd}$ order perturbation. Other parameters:
$\tau_d/\tau_s=50$, $a = 0.5$ and $x \in [-\pi,\pi)$.\label{slow-decaying-behavior}}
\end{center}\end{figure}

The network dynamics displays a very interesting
behavior in the marginally unstable region of the
static bump. In this regime, the static bump
solution barely loses its stability.
The bump is stable if the level of synaptic depression is
low, but unstable at high
levels.  Since the STD time scale is much longer
than the synaptic time scale, a bump can exist before the
synaptic depression becomes effective.  This maintains
the bump in the plateau state with a slowly decaying amplitude,
as shown in Fig. \ref{slow-decaying-behavior}(a).  After a
time duration of the order of $\tau_d$, the STD strength becomes sufficiently
significant, as shown in Fig. \ref{slow-decaying-behavior}(b),
and the bump
state eventually decays to the silent state.

\subsection{First Order: Trajectory Analysis}

\begin{figure}[ht]\begin{center}
\includegraphics[width=8.3cm]{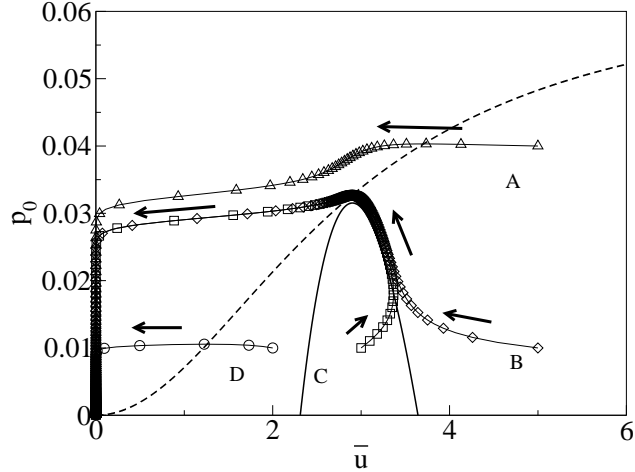}
\caption{ Trajectories of network dynamics starting from various initial conditions at
$(\overline{k}, \overline{\beta})$ = (0.95, 0.0085) (point P in Fig. \ref{phase-diagram-STD}). Solid line: $\overline{u}$-nullcline. Dashed
line: $p_0$-nullcline. Symbols are data points spaced at time intervals of $2\tau_s$.  Parameter: $\tau_d/\tau_s = 50$.
\label{trajectory-theory}}
\end{center}\end{figure}

It is instructive to analyze the plateau behavior first by using the first-order perturbation. We select a point in the marginally unstable
regime of the silent phase, that is, in the vicinity of the static phase. As shown in Fig. \ref{trajectory-theory}, the nullclines of
$\overline{u}$ and $p_{0}$ ($d\overline{u}/dt=0$ and $dp_{0}/dt=0$ respectively) do not have any intersections as they do in the static phase
where the bump state exists. Yet, they are still close enough to create a region with very slow dynamics near the apex of the $u$-nullcline at
$(\overline{u},p_{0})=[(8/\overline{k})^{1/2}, \sqrt{7/4}(1-\sqrt{\overline{k}})]$. Then, in Fig. \ref{trajectory-theory}, we plot the
trajectories of the dynamics starting from different initial conditions.

The most interesting family of trajectories is
represented by B and C in
Fig. \ref{trajectory-theory}. Due to the much
faster dynamics of $\overline{u}$, trajectories
starting from a wide range of initial conditions
converge rapidly, in a time of the order of
$\tau_{s}$, to a common trajectory in the vicinity of the $\overline{u}$-nullcline.
Along this common trajectory, $\overline{u}$ is
effectively the steady state solution of Eq.
(\ref{eq:dU0_app}) at the instantaneous value of
$p_{0}(t)$, which evolves with the much longer
time scale of $\tau_{d}$. This gives rise to the
plateau region of $\overline{u}$ which can
survive for a duration of the order of $\tau_{d}$.
The plateau ends after the trajectory has passed
the slow region near the apex of the
$\overline{u}$-nullcline. This dynamics is in
clear contrast with trajectory D, in which the
bump height decays to zero in a time of the order of
$\tau_s$.

Trajectory A represents another family of trajectories
having rather similar behaviors, although the lifetimes
of their plateaus are not so
long. These trajectories start from more depleted initial
conditions, and hence do not have the chance to get close to the
$\overline{u}$-nullcline. Nevertheless, they converge
rapidly, in a time of the order of $\tau_{s}$, to the band
$\overline{u}\approx(8/\overline{k})^{1/2}$, where the
dynamics of $\overline{u}$ is slow. The trajectories
then rely mainly on the dynamics of
$p_{0}$ to carry them out of this slow region, and
hence plateaus with lifetimes of the order of $\tau_{d}$ are created.


Following similar arguments, the plateau behavior also
exists in the stable region of the static
states. This happens when the initial
condition of the network lies outside the basin of
attraction of the static states, but still in
the vicinity of the basin boundary.

\begin{figure}[ht]\begin{center}
\includegraphics[width=8.3cm]{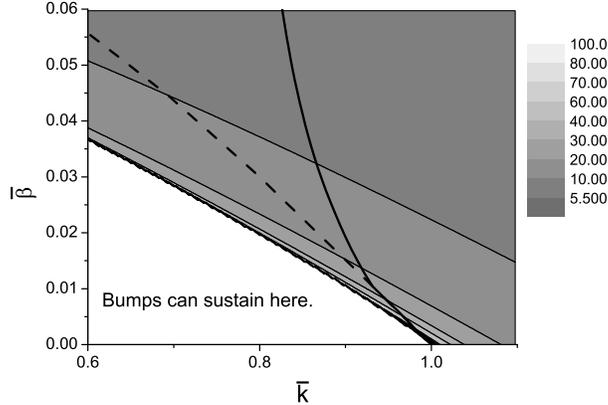}
\caption{ Contours of plateau lifetimes in the space of $\overline{k}$ and $\overline{\beta}$. The lines
are the two topmost phase boundaries in Fig. \ref{phase-diagram-STD}. In the initial condition, $\overline{A} = 4.82843$.\label{life-span}}

\end{center}\end{figure}

When one goes deeper into the silent phase, the
region of slow dynamics between the
$\overline{u}$- and $p_0$-nullclines broadens.
Hence plateau lifetimes are longest near the
phase boundary between the bump and silent
states, and become shorter when one goes deeper
into the silent phase. This is confirmed by the
contours of plateau lifetimes in the phase
diagram shown in Fig. \ref{life-span} obtained
numerically. The initial condition is
uniformly set by introducing an external stimulus
$I^{ext}(x|z_0)=A\exp[-x^2/(4a^2)]$ to the right
hand side of Eq. (\ref{eq:dyn}), where $A$ is the
stimulus strength. After the network has reached
a steady state, the stimulus is removed at $t=0$,
leaving the network to relax.  It is observed in Fig. \ref{life-span}
that the plateau behavior can be found in an
extensive region of the parameter space.

\subsection{Second Order: Lifetime Analysis}

As shown in Fig. \ref{slow-decaying-behavior},
the first-order perturbation over-estimates the
stability of the plateau state, yielding
lifetimes longer than the simulation results. The
main reason is that the width of the synaptic
depression profile is constrained to be a constant
in the first-order perturbation.  However, the
synaptic depression profile is broader than the
bump.  This can be seen from Eq. (\ref{eq:dp}),
rewritten as
\begin{equation}
  \tau_d\frac{\partial p(x,t)}{\partial t} =
  \left[1+\tau_d\beta r\left(x,t\right)\right]\left[1-p(x,t)-\frac{\tau_d\beta r(x,t)}{1+\tau_d \beta r(x,t)}\right],
\end{equation}
This shows
that the neurotransmitter loss, $1-p(x,t)$,
relaxes towards an expression consisting of the Gaussian $r(x,t)$,
normalized by the factor $1+\tau_d\beta r(x,t)$.
This normalization factor is smaller where the firing rate is low, so that
the profile of $1-p(x,t)$ is broader than the firing rate profile $r(x,t)$.

To incorporate the effects of a broadened STD profile, we introduce the second-order perturbation.  Dynamical equations are obtained by
truncating the equations beyond the second order. As shown in Fig. \ref{slow-decaying-behavior}, the second-order perturbation yields a much
more satisfactory agreement with simulation results than do lower order perturbations.

\section{Decoding with Enhanced Accuracy in CANNs with STF}
CANNs have been interpreted as an efficient framework for neural systems implementing
population decoding \cite{Deneve99,Wu02}. Consider the reading-out of an external feature $z_0$ from noisy inputs by CANNs. For example, $z_0$
may represent the moving direction of an object. In the decoding paradigm, a CANN responds to an external input $I^\text{ext}(x)$ with a bump
state $r(x|\hat{z})$, where the peak position of the bump $\hat{z}$ is interpreted as the decoding result of the network.

In the presence of STF, neuronal connections are
facilitated around the area where neurons are
most active. With this additional feature, the
network decoding will be determined not only by
the instantaneous input, but also by the recent history
of external inputs. Consequently, temporal
fluctuations in external inputs are largely
averaged out, leading to improved decoding
accuracies.

We consider an external input given by
\begin{equation}
I^\text{ext}(x,t)=A\exp\left[-\frac{\left(x-z_0-\eta(t)\right)}{4a^{2}}^{2}\right],
\end{equation}
where $z_0$ represents the true stimulus value,
and $\eta(t)$ white noise of zero mean
and satisfies $\left\langle
\eta(t)\eta(t')\right\rangle =2Ta^2\tau_s\delta(t-t')$
with $T$ denoting the noise strength.

\begin{figure}[ht]\begin{center}
\includegraphics[width=8.3cm]{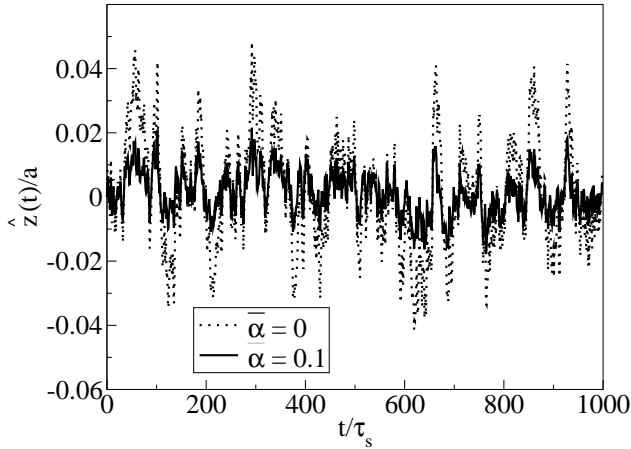}
 \caption{The typical decoding performance of the network with and without STF.
Parameters: $N=80$, $a = 0.5$, $x \in [-\pi, \pi)$ $\overline{k}=0.25$, $\overline{A}=1.596$ and $T = 0.02$.\label{decoding-performance}}

\end{center}\end{figure}

In the presence of weak noise, the position of the bump state is found to be centered at $z_0 + s(t)$, where $s(t)$ is the deviation of the center
of mass of the bump from the position of stimulus $z_0$, as derived in Appendix E.
Hence, the decoding error of the network is measured by the variance of the bump position over time, namely, $\langle s(t)^2\rangle$.
Fig. \ref{decoding-performance} shows the typical decoding performance of the network with and without STF. We see that with STF, the fluctuation
of the bump position is reduced significantly. Fig. \ref{decoding-result} compares the theoretical and measured decoding errors
in different noise strengths (see Appendix E).

\begin{figure}[ht]\begin{center}
\includegraphics[width=8.3cm]{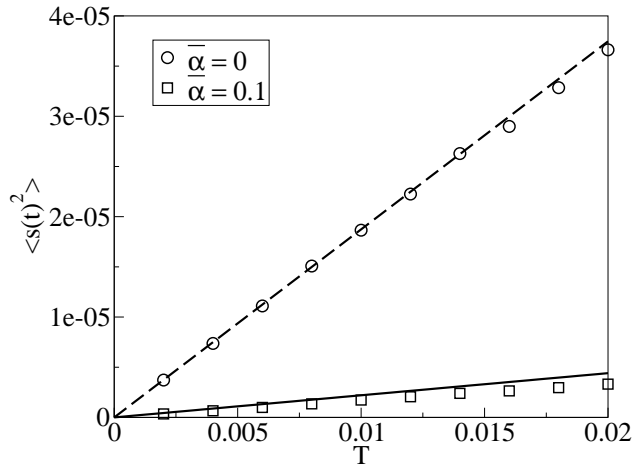}
\caption{The decoding errors of the network vs. different levels of noise. Parameters other than
$T$ are the same as those in Fig. \ref{decoding-performance}.  Symbols: simulations.  Dashed line: predictions for
$\overline{\alpha} = 0$.  Solid line: predictions for $\overline{\alpha} = 0.1$.\label{decoding-result}}

\end{center}\end{figure}

\section{Tracking with Enhanced Mobility in CANNs with STD}

A key property of CANNs is their capacity to track time-varying stimuli, which lays the foundation for CANNs to implement spatial navigation,
population decoding, and to update head-direction memory. To investigate the
tracking performance of CANNs, we consider
\begin{equation}
I^\text{ext}=A\exp\left[-\frac{\left(x-z_0\left(t\right)\right)^2}{4a^2}\right],
\end{equation}
where the stimulus position $z_0\left(t\right)$ is time dependent.

\begin{figure}[ht]\begin{center}
\includegraphics[width=8.3cm]{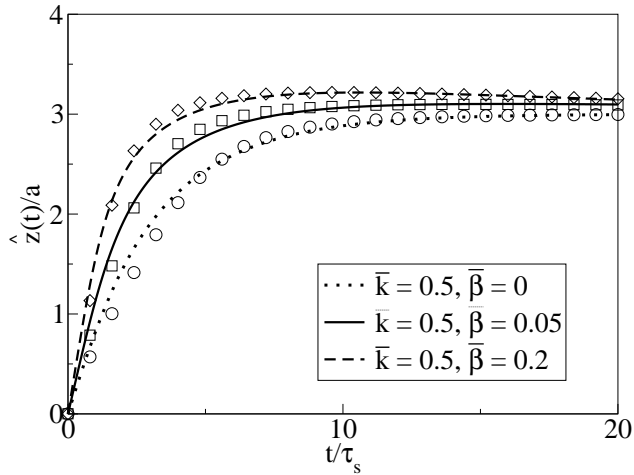}
\caption{ The response of CANNs with STD to a stimulus that changed abruptly from $z_0/a=0$ to
$z_0/a=3.0$ at $t=0$. Symbols: numerical solutions. Lines: Gaussian approximation using $11^{\rm th}$ order perturbation of the STD
coefficient. Parameters: $\tau_d/\tau_s=50$, $\overline{A} = 4.82843$, $N = 80$, $a = 0.5$ and $x \in
[-\pi,\pi)$.\label{tracking-performance-STD}}
\end{center}\end{figure}

We first investigate the impact of STD, and consider a tracking task in
which the $z_0\left(t\right)$ abruptly changes from 0 at $t=0$ to a new value at $t=0$.
Fig. \ref{tracking-performance-STD} shows the network responses during the tracking process. Compared with the case without STD, we find that
the bump shifts to the new position faster. When $\overline{\beta}$ is too strong, the bump may overshoot the target before eventually
approaching it. As remarked previously, this is due to the metastatic behavior of the bumps, which enhances their readiness to move from the
static state when a small push is exerted.

\begin{figure}[ht]\begin{center}
\includegraphics[width=8.3cm]{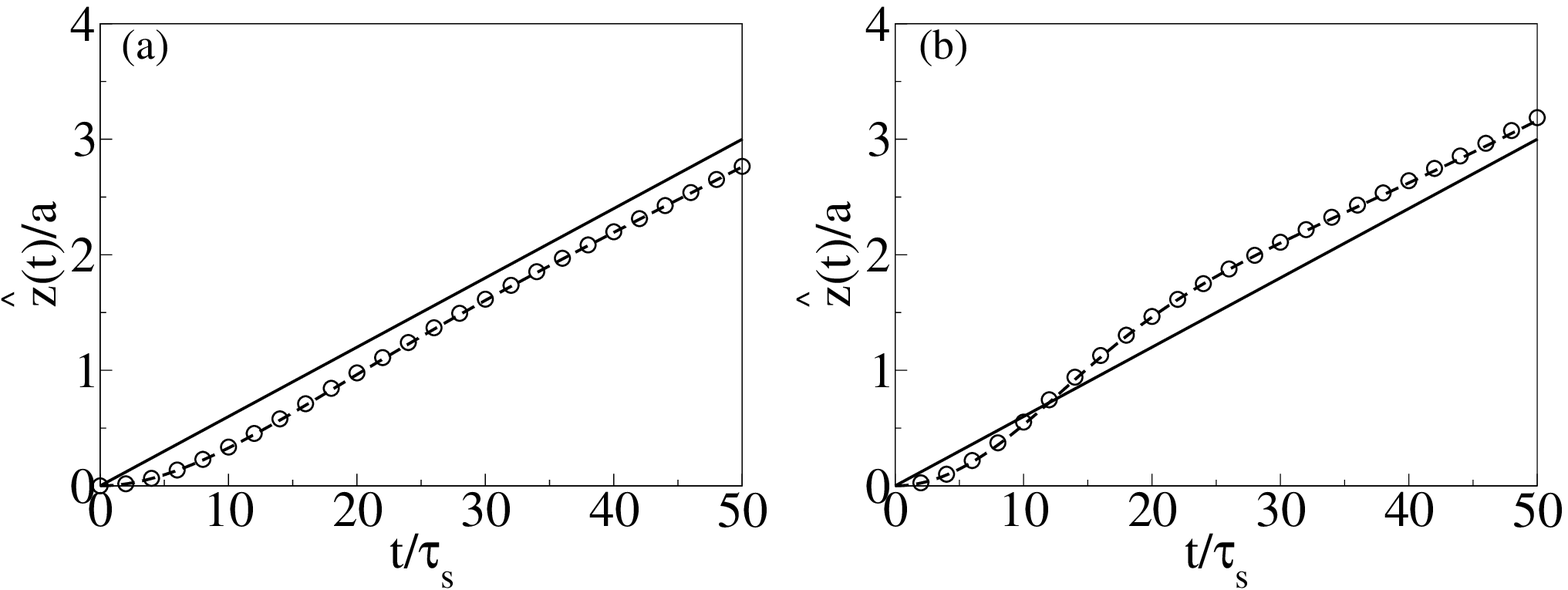}
 \caption{Tracking of a neuronal bump with
a continuously moving stimulus. Symbols: the peak of the bump from simulation. Dashed line: $11^\text{th}$ order perturbation prediction.
Solid line: a continuously moving stimulus with speed $\tau_sv/a =  0.06$. Parameters: $\overline{A} = 1.5958$, $\overline{k} = 0.5$. (a)
$\overline{\beta} = 0.01$. (b) $\overline{\beta} = 0.05$. Other parameters are the same as those in Fig. \ref{tracking-performance-STD}.
\label{tracking-lagging-STD}}
\end{center}\end{figure}

We also study the tracking of an external stimulus moving with a constant velocity $v$, that is, $z_0\left(t\right)$
changes from 0 to $vt$ at $t=0$. As shown in Fig. \ref{tracking-lagging-STD}(a), when STD is weak,
the initial speed of the bump is almost zero. Then, when the stimulus is moving away, the bump accelerates in an attempt
to catch up with the stimulus.  After some time, the separation between the bump and the stimulus converges to a constant.  This tracking
behavior is similar to the case without STD. The tracking behavior in the case of strong STD is more interesting. As shown in
Fig. \ref{tracking-lagging-STD}(b), the bump position eventually overtakes the stimulus, displaying an anticipative behavior.  This can be
attributed to the metastatic property of STD.

\begin{figure}\begin{center}
\includegraphics[width=8.3cm]{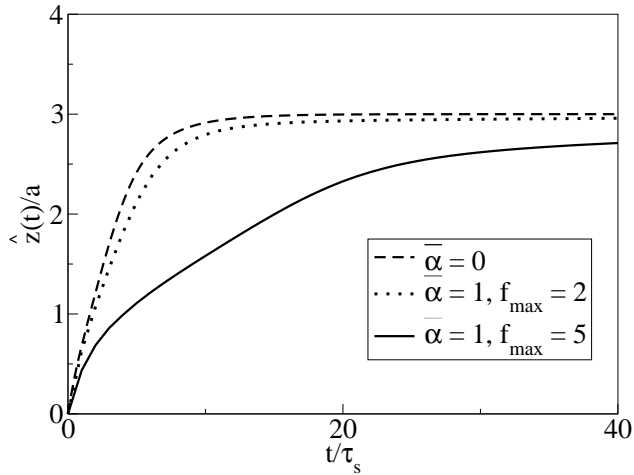}
 \caption{The response of CANNs with STD to a stimulus that changed abruptly from $z_0/a=0$ to
$z_0/a=3.0$ at $t=0$.  Parameters: $\overline{A} = 4.82843$, $\tau_d/\tau_s = 50$ and $\tau_f/\tau_s = 50$. \label{tracking-performance-STF}}
\end{center}\end{figure}


We further explore how STF affects the tracking performance of CANNs.
In general, there is a trade-off between the stability of network states
and the capacity of the network to track time-varying stimuli.  Since
STD and STF have opposite effects on the mobility of the network states,
we expect that they will also have opposite impacts on the
tracking performance of CANNs.  Indeed,
STF degrades the tracking performance of CANNs (see
Fig. \ref{tracking-performance-STF}). The larger the STF strength, the slower the tracking speed of the network.

\section{Conclusions and Discussions}

In the present study, we have investigated the impact of STD and STF on the dynamics of CANNs and their potential roles in neural information
processing. We have analyzed the dynamics using successive orders of perturbation. The perturbation analysis works well when STD is not too
strong, although it over-estimates the stability of the bumps when STD is strong. The zeroth order analysis accounts for the Gaussian shape of
the bump, and hence can predict the boundary of the static phase satisfactory. The first-order analysis includes the displacement mode and
asymmetry with respect to the bump peak, and hence can describe the onset of the moving phase. Furthermore, it provides insights on the
metastatic nature of the bumps and its relation with the enhanced tracking performance.  The second-order analysis further includes the width
distortions, and hence improves the prediction of the boundary of the moving phase, as well as the lifetimes of the plateau states.
Higher-order perturbations are required to yield more accurate descriptions of results such as the overshooting in the tracking process.  We anticipate that
the perturbation analysis will also be useful in many other
population decoding problems, such as in quantifying the deformation
of tuning curves due to neural adaptation \cite{Cortes11}.

More importantly, our work reveals a number of
interesting behaviors which may have far-reaching
implications in neural computation.

First, STD endows CANNs with slow-decaying behaviors. When a network is initially stimulated to an active state by an external input, it will
decay to the silent state very slowly after the input is removed. The duration of the plateau is of the time scale of STD rather than of neural signaling.
This provides a way for the network to hold the stimulus information for up to hundreds of milliseconds, if the network operates in the
parameter regime where the bumps are marginally unstable. This property is, on the other hand, extremely difficult to implement in
attractor networks without STD. In a CANN without STD, an active state of the network will either decay to the silent state exponentially fast or be
retained forever, depending on the initial activity level of the network. Indeed, how to shut off the activity of a CANN gracefully has been a challenging
issue that has received wide attention in theoretical neuroscience, with researchers suggesting that a strong external input either in the form of
inhibition or excitation must be applied (see, e.g., Gutkin, Laing, Colby, Chow, and Ermentrout (2001)). Here, we have shown that in certain circumstances, STD can provide a mechanism
for closing down network activities naturally and after a desirable duration. Taking into account the time scale of STD (in the order of $100$ ms)
and the passive nature of its dynamics, the STD-based memory is most likely associated with the sensory memory of the brain, e.g., the
iconic and the echoic memories \cite{Baddeley99}.

Second, with STD, CANNs can support both static and moving bumps. Static bumps exist only when the synaptic depression is sufficiently weak. A
consequence of synaptic depression is that static bumps are placed in the metastatic state, so that its response to changing stimuli is speeded
up, enhancing its tracking performance. The states of moving bumps may be associated with the travelling wave behaviors widely observed in the
neurocortex. We have also observed that for strong STD, the network state can even overtake the moving stimulus, reminiscent of the anticipative
responses of head-direction and place cells~\cite{Blair95,Okeefe93}. It is interesting to note that this occurs in the parameter range where the
network holds spontaneous moving bump solutions, suggesting that travelling wave phenomena may be closely related to the predicting capacity of
neural systems.

Third, STF improves the decoding accuracy of CANNs.
When an external stimulus is presented, STF strengthens
the interactions among neurons
which are tuned to the stimulus. This stimulus-specific
facilitation provides a mechanism for the network to hold a memory trace of external
inputs up to the time scale of STF, and this
information can be used by the neural system for
executing various memory-based operations, such
as operating the working memory. We have tested this idea in a
population decoding task, and found that the
error is indeed decreased. This is due to
the determination of the
network response by both the instantaneous value
and the history of external inputs, which effectively
averages out temporal
fluctuations.

These computational advantages of dynamical synapses lead to the following implications for the modeling of neural systems. First, it sheds
some light on the long-standing debate in the field about the instability of CANNs in the presence of noise. Two aspects of instability have been identified
\cite{Wu05,Seung00}. One is the structural instability, which refers to the argument that network components in reality, such as the neuronal
synapses, are unlikely to be as perfect as mathematically required in CANNs. A small amount of discrepancy in the network structure can destroy
the state space considerably, destabilizing the bump state after the stimulus is removed. The other instability refers to the computational
sensitivity of the network to input noises. Because of neutral stability, the bump position is very susceptible to fluctuations in external
inputs, rendering the network decoding unreliable. We have shown that STF can largely improve the computational robustness of CANNs by
averaging out the temporal fluctuations in inputs. Similarly, STF can overcome the structural susceptibility of CANNs. With STF, the neuronal
connections around the bump area are strengthened temporally, which effectively stabilizes the bump on the time scale of STF \cite{Tsodyks11}.
Another mechanism in a similar spirit is the reduction of the inhibition strength around the bump area \cite{Cater07}.

Second, STD and STF should be dominant in different
areas of the brain.  We have investigated the impact of STD and
STF on the tracking performance of CANNs.
There is, in general, a trade-off between the
stability of bump states and the tracking
performance of the network. STD increases
the mobility of bump states and hence the tracking speed of
the network, whereas STF has the opposite
effect. These differences predict that in
cortical areas where time-varying stimuli, such as the
head direction and the moving direction of objects,
are encoded, STD should have a
stronger effect than STF. On the other hand, in cortical
areas where the robustness of bump states, i.e.,
the decoding accuracy of stimuli, is preferred,
STF should have a stronger effect.

Third, STD and STF consume different levels of energy and operate on different time scales.
We have shown that both STD and STF can generate temporal memories, but their ways of achieving it are quite different. In STD, the
memory is held in the prolonged neural activities, whereas in STF, it is in the facilitated neuronal connections.
Mongillo, Barak, and Tsodyks (2008) proposed that with STF, neurons may not even have to be active after the stimulus is removed. The facilitated neuronal
connections, mediated by the elevated calcium residue, is sufficient to carry out the memory retrieval. In our model, this is equivalent to
setting the network in the parameter regime without static bump solutions, or in the regime with static bump solutions but with the external
stimulus presented for such a short time that neuronal interactions cannot be fully facilitated. Thus, taking into account the energy consumption associated
with neural firing, the STF-based mechanism for short-term memory has the advantage of being economically efficient. However, the STD-based one
also has the desirable property of enabling the stimulus information to be propagated to other cortical areas, since neural firing is necessary
for signal transmission, and this is critical in the early information pathways. Furthermore, the time durations required for eliciting STF-
and STD-based memory are significantly different. The former needs a stimulus to be presented for an amount of time up to $\tau_f$ to
facilitate neuronal interactions sufficiently; whereas the latter simply requires a transient appearance of a stimulus. This difference
implies that the two memory mechanisms may have potentially different applications in neural systems.

In summary, we have revealed that STP can play very valuable roles in neural information processing, including achieving temporal memory,
improving decoding accuracy, enhancing tracking performance and stabilizing CANNs. We have also shown that STD and STF tend to have different
impacts on the network dynamics. These results, together with the fact that STP displays large diversity in the neural cortex, suggest that the
brain may employ a strategy of weighting STD and STF differentially depending on the computational task. In the present study, for the
simplicity of analysis, we have only explored the effects of STD and STP separately. In practice, a proper combination of STD and STP can make
the network exhibit new and interesting behaviors and implement new and computationally desirable properties. For instance, a CANN with both STD and
STF, and with the time scale of the former shorter than that of the latter, can hold bump states for a period of time before shifting the
memory to facilitated neural connections. This enables the network to achieve both goals of conveying the stimulus information to other
cortical areas and holding the memory cheaply. Alternatively, the network may have the time scale of STD longer than that of STF, so that the
network can produce improved encoding results for external stimuli and also close down bump activities easily. We will explore these
interesting issues in the future.

\section*{Acknowledgment}
We acknowledge the valuable comments of Terry Sejnowski on this work. This study is supported by the Research Grants Council of Hong Kong (grant
numbers 604008 and 605010) and the 973 program of China (grant number 2011CBA00406).

\newpage

\renewcommand{\theequation}{A-\arabic{equation}}
\setcounter{equation}{0}

\section*{Appendix A: Consistency with the Model of Tsodyks et al}

Our modeling of the dynamics of STP is consistent with the phenomenological model proposed by Tsodyks, Pawelzik, and Markram (1998).
They modeled STD by considering $p$ as the fraction of utilizable neurotransmitters, and STF by introducing $U_0$ as the
release probability of the neurotransmitters. The release probability $U_0$ relaxes to a nonzero constant, $u_\text{rest}$, but is enhanced at
the arrival of a spike by an amount equal to $u_0(1-U_0)$. Hence the dynamics of $p$ and $U_0$ are given by
\begin{eqnarray}
  \tau_s \frac{\partial u}{\partial t}
  & = & I_\text{ext} - u + \rho\int dx' J(x-x')p(x')U_1(x')r(x'), \\
  \tau_d\frac{\partial p}{\partial t} & = & 1-p - U_1\tau_d pr, \\
  \tau_f\frac{\partial U_0}{\partial t} & = & u_\text{rest}-U_0 + u_0(1-U_0)\tau_f r,\label{eq:A-3}
\end{eqnarray}
where $U_1\equiv u_0+U_0(1-u_0)$ is the release probability of the neurotransmitters after the arrival of a spike. The $x$ and $t$
dependence of $u$, $p$, $r$ and $U_0$ are omitted in the above equations for convenience. Eliminating $U_0$, we obtain from Eq. \eqref{eq:A-3}
\begin{equation}
\frac{\partial U_1}{\partial t} 
= \frac{u_0+\left(1-u_0\right)u_\text{rest}-U_1}{\tau_f}
+u_0\left(1-U_1\right)r.
\end{equation}
Substituting
 $\alpha$, $\beta$ and
$f$ via $\alpha = u_0$, $\beta = u_0+(1-u_0)u_\text{rest}$, $U_1=[u_0+(1-u_0)u_\text{rest}](1+f)$, 
$f_\text{max} = (1-\beta)/\beta$, we obtain Eqs. \eqref{eq:dp} and \eqref{eq:df}.
Rescaling  $\beta J$ to $J$, we obtain Eq. (\ref{eq:dyn}). $\alpha$
and $\beta$ are the STF and STD parameters respectively, subject to $\beta \ge \alpha$.

\renewcommand{\theequation}{B-\arabic{equation}}
\setcounter{equation}{0}

\section*{Appendix B: The Perturbation Approach for Solving the Dynamics of CANNs with STD}

We use the perturbation approach to solve the network dynamics. Substituting Eqs. (\ref{eq:u_sum_u}) and (\ref{eq:p_sum_p}) into Eq.
(\ref{eq:dyn}), the right-hand side of Eq. (\ref{eq:dyn}) becomes
\begin{eqnarray}
  & & \frac{\rho}{B} \left[ \sum_{nm}a_na_m \int dx' J(x,x') v_n(x',t) v_m(x',t) \right.\nonumber \\
  & &  \left.- \sum_{nml}a_na_mb_l \int dx' J(x,x') v_n(x',t) v_m(x',t) w_l(x',t) \right] \nonumber \\
  & & -\sum_na_n(t)v_n(x,t) + I^\text{ext}(x,t).
\end{eqnarray}
This expression can be resolved into a linear combination of the distortion
modes $v_k(x,t)$.  The coefficients of these modes are obtained by multiplying
the expression with $v_k(x,t)$ and integrating $x$.  Using the orthonormal property
of the distortion modes, we have
\begin{equation}
  \sum_k v_k(x,t)\left[\frac{\rho C_r J_0}{B}\left(\sum_{nm} \mC^k_{nm}a_na_m-\sum_{nml}\mD^k_{nml}a_na_mb_l\right)
    -a_k+I_k\right],
\end{equation}
where
\begin{equation}
  \mC^k_{nm}\equiv \int dx'' \int dx' J(x'',x')v_k(x'',t)v_n(x',t)v_m(x',t),
\end{equation}
\begin{equation}
  \mD^k_{nml}\equiv \int dx'' \int dx' J(x'',x')v_k(x'',t)v_n(x',t)v_m(x',t)w_l(x',t),
\end{equation}
and $I_k(t)$ is the $k^{\rm th}$ component of $I^\text{ext}(x,t)$.

Similarly, the right-hand side of Eq. (\ref{eq:dp}) becomes
\begin{equation}
   - \sum_k  w_k(x,t) \left[-b_k + \frac{\tau_d \beta}{B}\left(\sum_{nm} \mE^k_{nm}a_na_m  - \sum_{nml} \mF^k_{nml}a_na_mb_l\right)\right],
\end{equation}
where
\begin{equation}
  \mE^k_{nm}\equiv \int dx' w_k(x',t)v_n(x',t)v_m(x',t),
\end{equation}
\begin{equation}
  \mF^k_{nml}\equiv \int dx' w_k(x',t)v_n(x',t)v_m(x',t)w_l(x',t).
\end{equation}
We choose $v_n(x,t) = v_n(x-z(t))$ and $w_n(x,t) = w_n(x-z(t))$.  Using the following relationship of Hermite polynomials
\begin{eqnarray}
  H_{n+1}(x) & = & 2xH_n(x) - 2n H_{n-1}(x), \label{eq:Hermite_1} \\
  H'_n(x) & = & 2nH_{n-1}(x), \label{eq:Hermite_2}
\end{eqnarray}
we have $\dot{v}_n = (\dot{z}/(2a)) (\sqrt{n} v_{n-1} - \sqrt{n+1} v_{n+1})$
and $\dot{w}_n = (\dot{z}/(\sqrt{2}a)) (\sqrt{n} w_{n-1} - \sqrt{n+1} w_{n+1})$.
Making use of the orthonormality of $v_n$'s and $w_n$'s, we have
\begin{eqnarray}
 \tau_s\left[\dot{a}_k - \frac{\dot{z}}{2a}\left(\sqrt{k+1}a_{k+1}-\sqrt{k}a_{k-1}\right)\right]
&=& -a_k + \frac{\rho C_r J_0}{B}\left(\sum_{nm} \mC^k_{nm}a_na_m - \sum_{nml} \mD^k_{nml}a_na_mb_l\right),  \nonumber \\ & & \\
\tau_d\left[\dot{b}_k - \frac{\dot{z}}{\sqrt{2}a}\left(\sqrt{k+1}b_{k+1}-\sqrt{k}b_{k-1}\right)\right]
  &=& -b_k + \frac{\tau_d \beta}{B}\left(\sum_{nm} \mE^k_{nm}a_na_m - \sum_{nml} \mF^k_{nml}a_na_mb_l\right). \nonumber \\ & &
\end{eqnarray}

The values of $\mC^k_{nm}$, $\mD^k_{nml}$, $\mE^k_{nm}$ and $\mF^k_{nml}$ can be obtained from recurrence relations derived using
integration by parts and the relationships (\ref{eq:Hermite_1}) and (\ref{eq:Hermite_2}), which are given by
\begin{equation}
  \mC^k_{nm} = \frac{1}{2}\left(\sqrt{\frac{n}{k}}\mC^{k_1}_{n_1m}+\sqrt{\frac{m}{k}}\mC^{k_1}_{nm_1}\right),
\end{equation}
\begin{equation}
  \mC^k_{nm} = -\frac{1}{4}\sqrt{\frac{n_1}{n}}\mC^{k}_{n_2m} + \frac{3}{4}\sqrt{\frac{m}{n}}\mC^{k}_{n_1m_1}
  + \frac{1}{2}\sqrt{\frac{k}{n}}\mC^{k_1}_{n_1m},
\end{equation}
\begin{equation}
  \mC^k_{nm} = -\frac{1}{4}\sqrt{\frac{m_1}{m}}\mC^{k}_{nm_2} + \frac{3}{4}\sqrt{\frac{n}{m}}\mC^{k}_{n_1m_1}
  + \frac{1}{2}\sqrt{\frac{k}{m}}\mC^{k_1}_{nm_1},
\end{equation}
where $n_1\equiv n-1$ and $m_2\equiv m-2$ etc. in the indices. Similarly,
\begin{eqnarray}
  \mD^k_{nml} & = & \frac{2}{7}\left(\sqrt{\frac{n}{k}}\mD^{k_1}_{n_1ml} + \sqrt{\frac{m}{k}}\mD^{k_1}_{nm_1l} + \sqrt{2}\sqrt{\frac{l}{k}}\mD^{k_1}_{nml_1}\right) -\frac{1}{7}\sqrt{\frac{k_1}{k}}\mD^{k_2}_{nml},\\
  \mD^k_{nml} & = & \frac{6}{7}\left(\sqrt{\frac{1}{2}}\sqrt{\frac{n}{l}}\mD^{k}_{n_1ml_1} + \sqrt{\frac{1}{2}}\sqrt{\frac{m}{l}}\mD^{k}_{nm_1l_1} + \sqrt{\frac{2}{3}}\sqrt{\frac{k}{l}}\mD^{k_1}_{nml_1}\right) - \frac{1}{7}\sqrt{\frac{l_1}{l}}\mD^{k}_{nml_2},\\
  \mD^k_{nml} & = & -\frac{4}{7}\sqrt{\frac{n_1}{n}}\mD^k_{n_2ml} + \frac{3}{7}\sqrt{\frac{m}{n}}\mD^k_{n_1m_1l}  + \frac{3\sqrt{2}}{7}\sqrt{\frac{l}{n}}\mD^k_{n_1ml_1} + \frac{2}{7}\sqrt{\frac{k}{n}}\mD^{k_1}_{n_1ml},\\
  \mD^k_{nml} & = & -\frac{4}{7}\sqrt{\frac{m_1}{m}}\mD^k_{nm_2l} + \frac{3}{7}\sqrt{\frac{n}{m}}\mD^k_{n_1m_1l}   + \frac{3\sqrt{2}}{7}\sqrt{\frac{l}{m}}\mD^k_{nm_1l_1} + \frac{2}{7}\sqrt{\frac{k}{m}}\mD^{k_1}_{nm_1l},\\
  \mE^k_{nm} & = & \frac{1}{\sqrt{2}}\left(\sqrt{\frac{n}{k}}\mE^{k_1}_{n_1m}+\sqrt{\frac{m}{k}}\mE^{k_1}_{nm_1}\right), \\
  \mE^k_{nm} & = & -\frac{1}{2}\sqrt{\frac{n_1}{n}}\mE^{k}_{n_2m} + \frac{1}{2}\sqrt{\frac{m}{n}}\mE^{k}_{n_1m_1}
  + \frac{1}{\sqrt{2}}\sqrt{\frac{k}{n}}\mE^{k_1}_{n_1m}, \\
  \mE^k_{nm} & = & -\frac{1}{2}\sqrt{\frac{m_1}{m}}\mE^{k}_{nm_2} + \frac{1}{2}\sqrt{\frac{n}{m}}\mE^{k}_{n_1m_1}
  + \frac{1}{\sqrt{2}}\sqrt{\frac{k}{m}}\mE^{k_1}_{nm_1} , \\
  \mF^k_{nml} & = & \frac{\sqrt{2}}{3}\left(\sqrt{\frac{n}{k}}\mF^{k_1}_{n_1ml} + \sqrt{\frac{m}{k}}\mF^{k_1}_{nm_1l}  + \sqrt{2}\sqrt{\frac{l}{k}}\mF^{k_1}_{nml_1}\right) -\frac{1}{3}\sqrt{\frac{k_1}{k}}\mF^{k_2}_{nml},\\
  \mF^k_{nml} & = & \frac{\sqrt{2}}{3}\left(\sqrt{\frac{n}{l}}\mF^{k}_{n_1ml_1} + \sqrt{\frac{m}{l}}\mF^{k}_{nm_1l_1} + \sqrt{2}\sqrt{\frac{k}{l}}\mF^{k_1}_{nml_1}\right) -\frac{1}{3}\sqrt{\frac{l_1}{l}}\mF^{k}_{nml_2},\\
  \mF^k_{nml} & = & -\frac{2}{3}\sqrt{\frac{n_1}{n}}\mF^k_{n_2ml} + \frac{1}{3}\sqrt{\frac{m}{n}}\mF^k_{n_1m_1l} + \frac{\sqrt{2}}{3}\left(\sqrt{\frac{l}{n}}\mF^k_{n_1ml_1} + \sqrt{\frac{k}{n}}\mF^{k_1}_{n_1ml}\right),\\
  \mF^k_{nml} & = & -\frac{2}{3}\sqrt{\frac{m_1}{m}}\mF^k_{nm_2l} + \frac{1}{3}\sqrt{\frac{n}{m}}\mF^k_{n_1m_1l} + \frac{\sqrt{2}}{3}\left(\sqrt{\frac{l}{m}}\mF^k_{nm_1l_1} + \sqrt{\frac{k}{m}}\mF^{k_1}_{nm_1l}\right).
\end{eqnarray}
Since $\mC^0_{00}$, $\mD^0_{000}$, $\mE^0_{00}$ and $\mF^0_{000}$ can be calculated explicitly, all other $\mC^k_{nm}$, $\mD^k_{nml}$,
$\mE^k_{nm}$ and $\mF^k_{nml}$ can be deduced.

Below, we analyze the dynamics of the bump in successive orders of perturbation, where the perturbation order is defined by the highest integer value of index $k$ involved in the approximation.  We start with the zeroth order perturbation to describe the behavior of the static bumps, since their profile is
effectively Gaussian.  We then move on to the first-order perturbation, which includes asymmetric distortions.  Since spontaneous movements of
the bumps are induced by asymmetric profiles of the synaptic depression, we demonstrate that the first-order perturbation is able to provide the
solution of the moving bump.  Proceeding to the second-order perturbation, we allow for the flexibility of varying the width of the bump and
demonstrate that this is important in explaining the lifetime of the plateau state.  Tracking behaviors are predicted by the 11$^\text{th}$ order perturbation.

\renewcommand{\theequation}{C-\arabic{equation}}
\setcounter{equation}{0}

\section*{Appendix C: Static Bump: Lowest-Order Perturbation}

Without loss of generality, we let $z=0$. Substituting Eqs. (\ref{eq:U0_app0}) and (\ref{eq:p_app0}) into Eq. (\ref{eq:dyn}), we get
\begin{eqnarray} \tau_{s}e^{-\frac{x^{2}}{4a^{2}}}\frac{du_{0}}{dt} & = &\frac{\rho C_r J_{0}u_{0}^{2}}{\sqrt{2}\left(1+\sqrt{2\pi}ak\rho
u_{0}^{2}\right)}\left[e^{-\frac{x^{2}}{4a^{2}}}-p_{0}\sqrt{\frac{2}{3}}e^{-\frac{x^{2}}{3a^{2}}}\right]
-u_{0}e^{-\frac{x^{2}}{4a^{2}}} +Ae^{-\frac{x^2}{4a^2}}.\end{eqnarray} Using the projection onto $v_0$, we find that
$\exp\left(-x^{2}/3a^{2}\right)\approx\sqrt{6/7}\exp\left(-x^{2}/4a^{2}\right)$. This reduces the equation to \begin{equation}
\tau_{s}\frac{du_{0}}{dt}=- u_0 + \frac{\rho C_rJ_{0}u_{0}^{2}}{\sqrt{2}\left(1+\sqrt{2\pi}ak\rho
u_{0}^{2}\right)}\left[1-\sqrt{\frac{4}{7}}p_{0}\right] + A.\end{equation} Introducing the rescaled variables $\overline{u}$ and
$\overline{k}$, we get Eq. (\ref{eq:dU0_app}).

Similarly, substituting Eqs. (\ref{eq:U0_app0}) and (\ref{eq:p_app0}) into Eq. (\ref{eq:dp}) gives \begin{equation}
\tau_{d}e^{-\frac{x^{2}}{2a^{2}}}\frac{dp_{0}}{dt}=-p_{0}e^{-\frac{x^{2}}{2a^{2}}}+\frac{\tau_{d}\beta u_{0}^{2}}{1+\sqrt{2\pi}ak\rho
u_{0}^{2}}\left(e^{-\frac{x^{2}}{2a^{2}}}-p_{0}e^{-\frac{x^{2}}{a^{2}}}\right).\end{equation} Making use of the projection
$\exp\left(-x^{2}/a^{2}\right)\approx\sqrt{2/3}\exp\left(-x^{2}/2a^{2}\right)$, the equation simplifies to \begin{equation}
\tau_{d}\frac{dp_{0}}{dt}=-p_{0}+\frac{\tau_{d}\beta u_{0}^{2}}{1+\sqrt{2\pi}ak\rho
u_{0}^{2}}\left(1-\sqrt{\frac{2}{3}}p_{0}\right).\end{equation} Rescaling the variables $u$, $k$, $\beta$ and $A$, we get Eq.
(\ref{eq:dp0_app}).

The steady state solution is obtained by setting the time derivatives in Eqs. (\ref{eq:dU0_app}) and (\ref{eq:dp0_app}) to zero, yielding
\begin{eqnarray}
\overline{u} & = & \frac{1}{\sqrt{2}}\frac{\overline{u}^{2}}{B}\left(1-\sqrt{\frac{4}{7}}p_{0}\right),\label{eq:sup:fixpt_u}\\
p_{0} & = & \frac{\overline{\beta}\overline{u}^{2}}{B}\left(1-\sqrt{\frac{2}{3}}p_{0}\right),\label{eq:sup:fixpt_p}\end{eqnarray} where
$B\equiv1+\overline{k}\overline{u}^{2}/8$ is the divisive inhibition.

\begin{figure}[t]
\centering
\includegraphics[width=8.3cm]{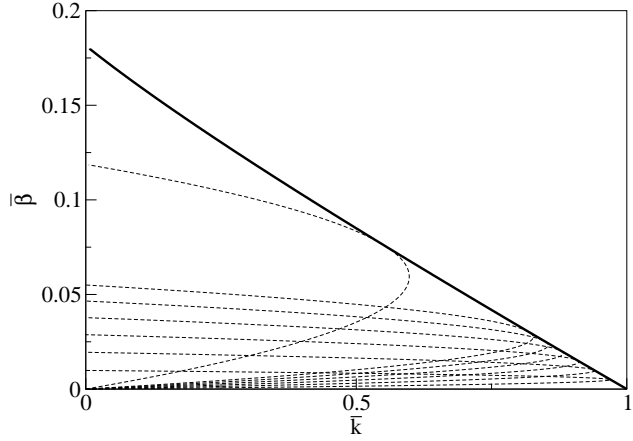}
\caption{\label{fig:fig1}The region of existence of static bump solutions. Solid line: the boundary of existence of static bump solutions.
Dashed lines: the parabolic curves for different constant values of $p_{0}$.}
\end{figure}

Dividing Eq. (\ref{eq:sup:fixpt_u}) by (\ref{eq:sup:fixpt_p}), we eliminate $B$ and get \begin{equation}
\overline{u}=\frac{1}{\sqrt{2}\,\overline{\beta}}\left(\frac{1-\sqrt{4/7}p_{0}}{1-\sqrt{2/3}p_{0}}\right)p_{0}.\end{equation} We can eliminate
$\overline{u}$ from Eq. (\ref{eq:sup:fixpt_p}). This gives rise to an equation for $p_{0}$
\begin{eqnarray}
    \frac{1}{2\overline{\beta}}\left(1-\sqrt{\frac{4}{7}}p_{0}\right)^{2}\left[1-\left(\sqrt{\frac{2}{3}}
    +\frac{\overline{k}}{8\overline{\beta}}\right)p_{0}\right]p_{0}
    -\left(1-\sqrt{\frac{2}{3}}p_{0}\right)^{2} = 0.
\end{eqnarray}
Rearranging the terms, we have \begin{equation}
\overline{k}=\frac{8}{p_{0}}\left(1-\sqrt{\frac{2}{3}}p_{0}\right)\overline{\beta}-\frac{16}{p_{0}^{2}}\left(\frac{1-\sqrt{2/3}p_{0}}{1-\sqrt{4/7}p_{0}}\right)^{2}\overline{\beta}^{2}.\label{eq:sup:exist_cond1}\end{equation}
Therefore, for each fixed $p_{0}$, we can plot a parabolic curve in the space of $\overline{\beta}$ versus $\overline{k}$. The dashed lines in
Fig.~\ref{fig:fig1} are parabolas for different values of $p_{0}$. The family of all parabolas map out the region of existence of static bumps.

\subsection*{Stability of the Static Bump}

\begin{figure}[t]
\centering \vspace{0.25cm}
\includegraphics[width=8.3cm]{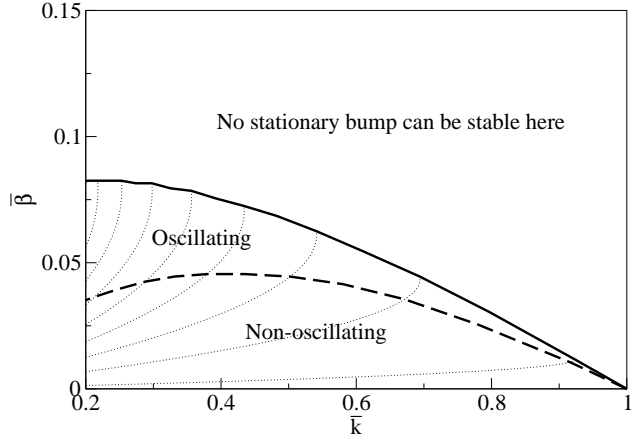}
\caption{\label{fig:fig2}The region of stable solutions of the static bump for $\tau_{d}/\tau_{s}=50$. Solid line: the boundary of stable
static bumps. Dashed line: the boundary separating the oscillating and non-oscillating convergence. Dotted lines: the curves for different
constant values of $p_{0}$.}
\end{figure}

To analyze the stability of the static bump, we consider the time evolution of $\epsilon=\overline{u}\left(t\right)-\overline{u}^{*}$ and
$\delta=p_{0}\left(t\right)-p_{0}^{*}$, where $\left(\overline{u}^{*},p_{0}^{*}\right)$ is the fixed point solution of Eqs.
(\ref{eq:sup:fixpt_u}) and (\ref{eq:sup:fixpt_p}). Then, we have
\begin{equation}
    \frac{d}{dt}\left(\begin{array}{c}
    \epsilon\\
    \delta\end{array}\right)
    = \left(\begin{array}{cc}
    A_{\epsilon\epsilon} & A_{\epsilon\delta} \\
    A_{\delta\epsilon} & A_{\delta\delta}\end{array}\right)
    \left(\begin{array}{c}\epsilon\\
    \delta\end{array}\right),
\end{equation}
where $A_{\epsilon\epsilon} =-1/\tau_s+\sqrt{2}\overline{u}(1-\sqrt{4/7}p_0)/\tau_sB^2$,
$A_{\epsilon\delta}=-\overline{u}^2\sqrt{2/3}/\tau_sB$, $A_{\delta\epsilon}=2\overline{\beta}\overline{u}(1-\sqrt{2/3}p_0)/\tau_dB^2$,
$A_{\delta\delta} =-1/\tau_d-\overline{\beta}\overline{u}^2\sqrt{2/3}/\tau_dB$. The stability condition is determined by the eigenvalues of the
stability matrix, $\left(T\pm\sqrt{T^{2}-4D}\right)/2$, where $D$ and $T$ are the determinant and the trace of the matrix respectively. Using
Eqs. (\ref{eq:sup:fixpt_u}) and (\ref{eq:sup:fixpt_p}), the determinant and the trace can be simplified to
\begin{equation}
    D=\frac{1}{\tau_{s}\tau_{d}B}
    \left(\frac{2\sqrt{4/7}p_{0}}{1-\sqrt{4/7}p_{0}}
    -\frac{2-B}{1-\sqrt{2/3}p_{0}}\right),
\end{equation}
\begin{equation}
    T=\frac{1}{\tau_{s}}\left[\frac{2}{B}-
    \frac{\tau_{s}}{\tau_{d}\left(1-\sqrt{2/3}p_{0}\right)}-1\right].
\end{equation}
The static bump is stable if the real parts of the eigenvalues are negative. The eigenvalues are real when $T^{2}\ge4D$. This corresponds to
non-oscillating solutions. After some algebra, we obtain the boundary $T^{2}=4D$ given by
\begin{eqnarray}
&&\left[\overline{\beta}-\frac{p_{0}\left(1-\sqrt{4/7}p_{0}\right)^{2}}{4\left(1-\sqrt{2/3}p_{0}\right)}+\frac{\tau_{s}}{\tau_{d}}\frac{p_{0}\left(1-\sqrt{4/7}p_{0}\right)^{2}}{4\left(1-\sqrt{2/3}p_{0}\right)^{2}}\right]^{2}=\frac{\tau_{s}}{\tau_{d}}\sqrt{\frac{4}{7}}\overline{\beta}\frac{p_{0}^{2}\left(1-\sqrt{4/7}p_{0}\right)}{1-\sqrt{2/3}p_{0}}.\nonumber \\\label{eq:sup:boundA14}
\end{eqnarray}
This boundary is shown in Fig.~\ref{fig:fig2}. Below this boundary, the stability condition can be obtained as
\begin{equation}
\overline{\beta}\le\frac{p_{0}\left(1-\sqrt{4/7}p_{0}\right)^{3}}{4\left(1-\sqrt{2/3}p_{0}\right)\left(1-2\sqrt{4/7}p_{0}+\sqrt{8/21}p_{0}^{2}\right)}.
\end{equation}
This upper bound is identical to the existence condition of (\ref{eq:sup:exist_cond1}), which is above the boundary of non-oscillating
solutions. This implies that all non-oscillating solutions are stable.

Above the boundary (\ref{eq:sup:boundA14}), the convergence to the steady state becomes oscillating, and the stability condition reduces to
$T\le0$, yielding Eq. (\ref{eq:static}). This condition narrows the region of static bump considerably, as shown in Fig. \ref{fig:fig2}.

\renewcommand{\theequation}{D-\arabic{equation}}
\setcounter{equation}{0}

\section*{Appendix D: Moving Bump: Lowest-Order Perturbation}

We substitute Eqs. (\ref{eq:moving_ansatz1}) and (\ref{eq:moving_ansatz2}) into Eqs. (\ref{eq:dyn}) and (\ref{eq:dp}). Eq. (1) becomes an
equation containing $\exp\left[-\left(x-vt\right)^{2}/4a^{2}\right]$ and $\exp\left[-\left(x-vt\right)^{2}/4a^{2}\right]\left(x-vt\right)/a$,
after making use of the projections $\exp[-(x-vt)^2/3a^2] \approx \sqrt{6/7}\exp[-(x-vt)^2/4a^2]$, and $\exp[-(x-vt)^2/3a^2] (x-vt)/a
\approx (\sqrt{6/7})^3\exp[-(x-vt)^2/4a^2] (x-vt)/a$.

Equating the coefficients of $\exp\left[-\left(x-vt\right)^{2}/4a^{2}\right]$ and
$\exp\left[-\left(x-vt\right)^{2}/4a^{2}\right]\left(x-vt\right)/a$, and rescaling the variables, we arrive at
\begin{equation}
    \overline{u}
    = \frac{\overline{u}^{2}}{\sqrt{2}B}
    \left(1-\sqrt{\frac{4}{7}}p_{0}\right),\quad
    \frac{v\tau_{0}}{2a}
    = \frac{\overline{u}}{B}\left(\frac{2}{7}\right)^{\frac{3}{2}}p_{1}.
\end{equation}
Similarly, making use of the projections $\exp[-(x-vt)^2/a^2] \approx \sqrt{2/3}\exp[-(x-vt)^2/2a^2]$, $\exp[-(x-vt)^2/a^2] (x-vt)/a \approx
(\sqrt{2/3})^3\exp[-(x-vt)^2/2a^2] (x-vt)/a$, we find that Eq. (\ref{eq:dp}) gives rise to
\begin{eqnarray}
    -\frac{v\tau_{d}}{2a}p_{1}
    & = & p_{0}-\frac{\overline{\beta}\overline{u}^{2}}{B}
    \left(1-\sqrt{\frac{2}{3}}p_{0}\right),\\
    \frac{v\tau_{d}}{a}p_{0}
    & = & \left[1+\frac{\overline{\beta}\overline{u}^{2}}{B}
    \left(\frac{2}{3}\right)^{\frac{3}{2}}\right]p_{1}.
\end{eqnarray}
The solution can be parametrized by $\xi\equiv\overline{\beta}\overline{u}^{2}/B$,
\begin{equation}
    p_{0}=\frac{\frac{\tau_{s}}{\tau_{d}}
    \left[1+\left(\frac{2}{3}\right)^{\frac{3}{2}}\xi\right]}{G(\xi)},
    \quad
    \frac{\overline{u}}{B}
    =\sqrt{2}\left(\frac{7}{4}\right)^\frac{3}{2}G(\xi),\nonumber\\
\end{equation}
\begin{equation}
    \frac{v\tau_{s}}{a}=\sqrt{2\frac{\tau_s}{\tau_d}F(\xi)},
    \quad
    p_1=\frac{\sqrt{4\frac{\tau_s}{\tau_d}F(\xi)}}{G(\xi)},
\label{eq:sup:A25}
\end{equation}
where $F(\xi)=(4/7)^{3/2}\xi-(\tau_s/\tau_d)[1+(2/3)^{3/2}\xi] [1-(\sqrt{2/3}-\sqrt{4/7})\xi]$, and
$G(\xi)=(4/7)^{3/2}+(4/7)^{1/2}(\tau_s/\tau_d)[1+(2/3)^{3/2}\xi]$. A real solution exists only if Eq. (\ref{eq:static2}) is satisfied. The
solution enables us to plot the contours of constant $\xi$ in the space of $\overline{k}$ and $\overline{\beta}$. Using the definition of
$\xi$, we can write
\begin{equation}
    \overline{k}
    =\frac{8}{\xi}\overline{\beta}-
    \frac{8}{\xi^{2}}\left(\frac{\overline{u}}{B}\right)^{2}
    \overline{\beta}^{2},
\end{equation}
where the quadratic coefficient can be obtained from Eq. (\ref{eq:sup:A25}). Fig. \ref{fig:fig3} shows the family of curves with constant
$\xi$, each with a constant bump velocity. The lowest curve saturates the inequality in Eq. (13), and yields the boundary between the static
and metastatic or moving regions in Fig. \ref{phase-diagram-STD}. Considering the stability condition in the next subsection, only the stable
branches of the parabolas are shown.

\subsection*{Stability of the Moving Bump}

To study the stability of the moving bump, we consider fluctuations around the moving bump solution. Suppose\begin{eqnarray}
u\left(x,t\right) & = & \left(u_{0}^{*}+u_{1}\right)e^{-\frac{\left(x-vt-s\right)^{2}}{4a^{2}}},\\
p\left(x,t\right) & = & 1-\left(p_{0}^{*}+\epsilon_{0}\right)e^{-\frac{\left(x-vt-s\right)^{2}}{2a^{2}}}
+\left(p_{1}^{*}+\epsilon_{1}\right)\left(\frac{x-vt-s}{a}\right)e^{-\frac{\left(x-vt-s\right)^{2}}{2a^{2}}}.\text{~~~}\end{eqnarray} These
expressions are substituted into the dynamical equations. The result is \begin{eqnarray}
\tau_{s}\frac{d\overline{u}_{1}}{dt} & = & \frac{\sqrt{2}\overline{u}}{B^{2}}\left(1-\sqrt{\frac{4}{7}}p_{0}\right)\overline{u}_{1}-\frac{\overline{u}^{2}}{B}\sqrt{\frac{2}{7}}\epsilon_{0}-\overline{u}_{1},\nonumber \\ \label{eq:sup:A31}\\
\frac{\tau_{s}}{a}\frac{ds}{dt} & = &
-\frac{v\tau_{s}}{a\overline{u}}\overline{u}_{1}+\frac{4p_{1}}{B^{2}}\left(\frac{2}{7}\right)^{\frac{3}{2}}\overline{u}_{1}+\frac{2\overline{u}}{B}\left(\frac{2}{7}\right)^{\frac{3}{2}}\epsilon_{1},\nonumber
\\ \label{eq:sup:A32}\end{eqnarray}\begin{eqnarray}
\tau_{d}\frac{d\epsilon_{0}}{dt} & = & \frac{2\overline{\beta}\overline{u}}{B^{2}}\left(1-\sqrt{\frac{2}{3}}p_{0}\right)\overline{u}_{1}-\left(1+\sqrt{\frac{2}{3}}\xi\right)\epsilon_{0}-\frac{v\tau_{d}}{2a}\epsilon_{1}-\frac{\tau_{d}p_{1}}{2a}\frac{ds}{dt},\label{eq:sup:A33}\\
\tau_{d}\frac{d\epsilon_{1}}{dt} & = & -\frac{2\overline{\beta}\overline{u}p_{1}}{B^{2}}\left(\frac{2}{3}\right)^{\frac{3}{2}}\overline{u}_{1}+\frac{v\tau_{d}}{a}\epsilon_{0}-\left[1+\left(\frac{2}{3}\right)^{\frac{3}{2}}\xi\right]\epsilon_{1}+\frac{\tau_{d}p_{0}}{a}\frac{ds}{dt}.\label{eq:sup:A34}\end{eqnarray}
We first revisit the stability of the static bump. By setting $v$ and
$p_{1}$ to 0, considering the asymmetric fluctuations $s$ and $\epsilon_{1}$ in Eqs. (\ref{eq:sup:A32}) and (\ref{eq:sup:A34}), and eliminating
$ds/dt$, we have
\begin{equation}
\tau_{s}\frac{d\epsilon_{1}}{dt}=\left\{
\frac{2\overline{u}}{B}\left(\frac{2}{7}\right)^{\frac{3}{2}}p_{0}-\frac{\tau_{s}}{\tau_{d}}\left[1+\left(\frac{2}{3}\right)^{\frac{3}{2}}\xi\right]\right\}
\epsilon_{1.}\end{equation} Hence the static bump remains stable when the coefficient of $\epsilon_1$ on the right-hand side is non-positive.
Using Eq. (\ref{eq:sup:A25}) to eliminate $p_{0}$ and $\overline{u}/B$, we recover the condition in Eq. (\ref{eq:static2}). This shows that the
bump becomes a moving one as soon as it becomes unstable against asymmetric fluctuations, as described in the main text.

\begin{figure}[t]
\centering\vspace{0.25cm}
\includegraphics[width=8.3cm]{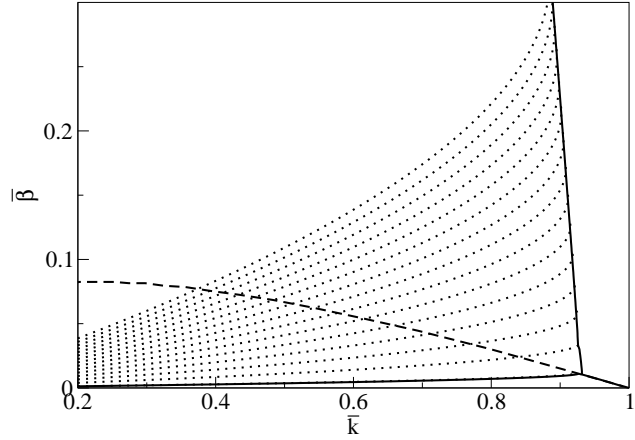}
\caption{\label{fig:fig3}The stable branches of the family of the curves with constant values of $\xi$ at $\tau_{d}/\tau_{s}=50$. The dashed
line is the phase boundary of the static bump.}
\end{figure}

Now we consider the stability of the moving bump. Eliminating $ds/dt$ and summarizing the equations into matrix form,
\begin{equation}
    \tau_{s}\frac{d}{dt}\left(\begin{array}{c}
    \overline{u}_{1}\\
    \epsilon_{0}\\
    \epsilon_{1}\end{array}\right)
    =\left(\begin{array}{ccc}
    \frac{2}{B}-1 & -\frac{\overline{u}^{2}}{B}\sqrt{\frac{2}{7}} & 0\\
    P_{0u} & P_{00} & -\frac{v\tau_{s}}{a}\\
    P_{1u} & \frac{v\tau_{s}}{a} & 0\end{array}\right)
    \left(\begin{array}{c}
    \overline{u}_{1}\\
    \epsilon_{0}\\
    \epsilon_{1}\end{array}\right),
\end{equation}
where $P_{0u}=2p_0\tau_s/B\overline{u}\tau_d+v\tau_s p_1/2a\overline{u}$,
$P_{00}=v\tau_sp_1/2ap_0-\overline\beta\overline{u}^2\tau_s/Bp_0\tau_d$, and $P_{1u}=2p_1\tau_s/B\overline{u}\tau_d-v\tau_sp_0/a\overline{u}$.
For the moving bump to be stable, the real parts of the eigenvalues of the stability matrix should be non-positive. The stable branches of the
family of curves are shown in Fig. \ref{fig:fig3}. The results show that the boundary of stability of the moving bumps is almost
indistinguishable from the envelope of the family of curves. Higher-order perturbations produce phase boundaries that agree more with
simulation results, as shown in Fig. \ref{phase-diagram-STD}.

\begin{figure}[t]
\centering
\includegraphics[width=8.3cm]{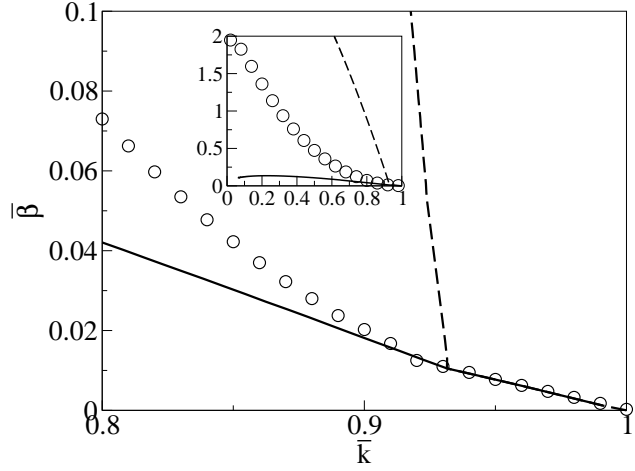}
\caption{\label{fig:fig4}The boundary of the moving phase. Symbols: simulation results. Dashed line: $1^\text{st}$ order perturbation. Solid
line: $2^\text{nd}$ order perturbation. Inset: The boundary of the moving phase in a broader range of $\beta$. Parameters: $N/L = 80/(2\pi)$,
$a/L = 0.5/(2\pi)$. }
\end{figure}

We compare the dynamical stability of the ansatz in Eqs. (\ref{eq:moving_ansatz1}) and (\ref{eq:moving_ansatz2}) with simulation results.  As
shown in Fig. \ref{fig:fig4}, the region of stability is over-estimated by the ansatz.  The major cause of this is that the width of the synaptic depression profile
is restricted to be $a$.  While this provides a self-consistent solution when STD is weak,  this
is no longer valid when STD is strong.  Due to the slow recovery of synaptic depression, its profile leaves a long, partially recovered tail
behind the moving bump, thus reducing the stability of the bump.  This requires us to consider the second-order perturbation, which takes into
account variation of the width of the STD profile.  As shown in Fig. \ref{fig:fig4}, the second-order perturbation yields a phase boundary
much closer to the simulation results when STD is weak.  However, as shown in the inset of Fig. \ref{fig:fig4}, the discrepancy increases
when STD is stronger, and higher-order corrections are required.

\renewcommand{\theequation}{E-\arabic{equation}}
\setcounter{equation}{0}

\section*{Appendix E: Decoding in CANNs with STF}

We start by considering bumps and STF profiles of the form
\begin{eqnarray}
\overline{u}(x,t) &=& \overline{u}_0 \exp \left[-\frac{(x-z_0-s(t))^2}{4a^2}\right], \label{eq:E1} \\
f(x,t) &=& f_0 \exp \left[-\frac{(x-z_0-s(t))^2}{2a^2}\right] + f_1 \left(\frac{x-z_0-s(t)}{a}\right)\exp \left[-\frac{(x-z_0-s(t))^2}{4a^2}\right].\label{eq:E2}
\end{eqnarray}
Note that since the noise occurs in the position of the bump, we can neglect changes in the height.
Substituting Eqs. (\ref{eq:E1}) and (\ref{eq:E1}) into Eq. (\ref{eq:dyn}), and
removing terms orthogonal to the position distortion mode, we have
\begin{equation}
\tau_{s}\frac{d}{dt}\left(\begin{array}{c}
\frac{s}{a}\\
\frac{f_{1}}{f_{0}}\end{array}\right)
 \equiv  M\left(\begin{array}{c}
\frac{s}{a}\\
\frac{f_{1}}{f_{0}}\end{array}\right)+\frac{\overline{A}}{\overline{u}_{0}a}\eta(t)\left(\begin{array}{c}
1\\
-1\end{array}\right),\end{equation} where
\begin{equation}
M=\left(\begin{array}{cc}
-\frac{\overline{A}}{\overline{u}_{0}} & \frac{2\overline{u}_{0}f_0}{B}\left(\frac{2}{7}\right)^{\frac{3}{2}}\\
\frac{\overline{A}}{\overline{u}_{0}} & -\left\{
\frac{2\overline{u}_{0}f_0}{B}\left(\frac{2}{7}\right)^{\frac{3}{2}}+\frac{\tau_{s}}{\tau_{f}}\left[1+\frac{\overline{\alpha}\overline{u}_{0}^{2}}{B}\left(\frac{2}{3}\right)^{\frac{3}{2}}\right]\right\}
\end{array}\right).
\end{equation}
This differential equation can be solved by first diagonalizing $M$. Let $-\lambda_\pm$ be the eigenvalues of $M$, and $(U_{s\pm}
~~U_{f\pm})^T$ be the corresponding eigenvectors.  Then the solution becomes
\begin{equation}
    \left(\begin{array}{c}
    \frac{s}{a}\\
    \frac{f_{1}}{f_{0}}\end{array}\right)=
    \frac{\overline{A}}{\overline{u}_{0}a}
    \int_{-\infty}^{t}\frac{dt_{1}}{\tau_s}\eta(t_{1})U
    \left(\begin{array}{cc}
    E_+ & 0\\
    0 & E_-\end{array}\right)
    U^{-1}\left(\begin{array}{c}
    1\\
    -1\end{array}\right),
\end{equation}
where $E_\pm=\exp[-\lambda_\pm(t-t_1)]$. Squaring the expression of $s/a$, averaging over noise, and integrating, we get
\begin{eqnarray}
    \left\langle \left(\frac{s}{a}\right)^2\right\rangle  & = &
    2T\left(\frac{\overline{A}}{\overline{u}_{0}}\right)^{2}
    \sum_{a,b=\pm}\left[U_{sa}\left(U_{as}^{-1}-U_{af}^{-1}\right)\right] \frac{1}{(\lambda_a+\lambda_b)\tau_s}
    \left[U_{sb}\left(U_{bs}^{-1}-U_{bf}^{-1}\right)\right].
\label{eq:decoding-error}
\end{eqnarray}



\end{document}